\documentclass[sigconf,nonacm]{acmart}
\usepackage[table]{xcolor}
\usepackage{hyperref}
\usepackage{booktabs}
\usepackage{multirow}
\usepackage{graphicx}
\usepackage{enumitem}
\usepackage{subfigure}  
\usepackage{graphicx}
\usepackage{bbding} 
\usepackage{stfloats}
\usepackage{float}
\usepackage{makecell}
\usepackage{algorithm} 
\usepackage{microtype}
\usepackage{algpseudocode}
\usepackage[title]{appendix}
\usepackage{bbding}
\usepackage{balance} 
\usepackage{xcolor}
\usepackage{tabularx}
\usepackage{array}
\usepackage{ragged2e}

\definecolor{headergray}{RGB}{245,245,245}
\definecolor{rowgray}{RGB}{240,240,240}
\definecolor{impgreen}{RGB}{0,100,0} 
\definecolor{mutedred}{RGB}{176,60,60}
\definecolor{lightgray}{gray}{0.95}

\setcopyright{none}
\begin{document}

\title{Interpretable Representation via LLM-Driven Generative Disentanglement for Local-Life Service Recommendation}

\author{Long Zhang}
\authornote{These authors contributed equally to this work.}
\affiliation{ %
  \institution{Kuaishou Technology}
  \city{Beijing}
  \country{China}}
\email{dragonzhang@mail.ustc.edu.cn}

\author{Hao Jiang}
\authornotemark[1]
\authornote{Corresponding author.}
\affiliation{ %
  \institution{Kuaishou Technology}
  \city{Beijing}
  \country{China}}
\email{jianghao11@kuaishou.com}

\author{Sheng Yu}
\affiliation{ %
  \institution{Kuaishou Technology}
  \city{Beijing}
  \country{China}}
\email{yusheng03@kuaishou.com}

\author{Fei Pan}
\affiliation{ %
  \institution{Kuaishou Technology}
  \city{Beijing}
  \country{China}}
\email{panfei05@kuaishou.com}

\author{Peng Jiang}
\affiliation{ %
  \institution{Kuaishou Technology}
  \city{Beijing}
  \country{China}}
\email{jiangpeng@kuaishou.com}

\author{Kun Gai}
\affiliation{%
  \institution{Kuaishou Technology}
  \city{Beijing}
  \country{China}}
\email{gai.kun@qq.com}

\begin{abstract}
While large language models (LLMs) have substantially advanced ID-based recommendation through Semantic ID (SID) modeling, many existing SID
generation frameworks still largely follow a
\textbf{single-representation-then-quantization} paradigm. 
This paradigm faces two key bottlenecks. First, \textbf{semantic entanglement} within the single representation mixes heterogeneous item attributes such as geography, brand, and category, causing key attributes to be insufficiently preserved during quantization, degrading SID quality, and leading to severe SID collisions. Second, \textbf{black-box} 
representation learning and quantization limit interpretability, because learned representations lack explicit attribute semantics and SID positions lack clear geographic or semantic correspondence.
To address these challenges, we propose
\textbf{\underline{I}}nterpretable \textbf{\underline{R}}epresentation via
LLM-Driven \textbf{\underline{G}}enerative
\textbf{\underline{D}}isentanglement for
\textbf{\underline{L}}ocal-Life Service Recommendation (LGRID). 
LGRID introduces a \textbf{Generative Disentanglement} paradigm, implemented through an SID construction pipeline of \emph{Encode} $\rightarrow$ \emph{Disentangle} $\rightarrow$ \emph{Align} $\rightarrow$ \emph{Quantize}.
It first encodes items with an LLM to preserve geographic--semantic dependencies, then routes the hidden states into attribute-aligned slots via a Structured Disentangled Block.
Synergistic Alignment Learning makes these slots decodable and discriminative. Dual-Stream Residual Quantization separately discretizes the geographic and semantic slots into compact SIDs.
Experiments on the Kuaishou and Foursquare datasets show that LGRID consistently outperforms strong SID baselines, achieving up to a 5.44\% relative AUC gain. 
LGRID also achieves over 99\% attribute-decoding accuracy for coarse-grained
geographic fields and reduces the full-SID collision rate to 39.9\%, compared with 97.0\% for LGSID. 
\end{abstract}

\ccsdesc[500]{Information systems~Recommender systems}
\vspace{-1.0cm}
\keywords{Local Services Recommendation; Large Language Model}
\maketitle

\vspace{-0.2cm}
\section{Introduction}

\begin{figure}[t]
  \centering  
\includegraphics[width=\linewidth,height=0.6\textheight,keepaspectratio]{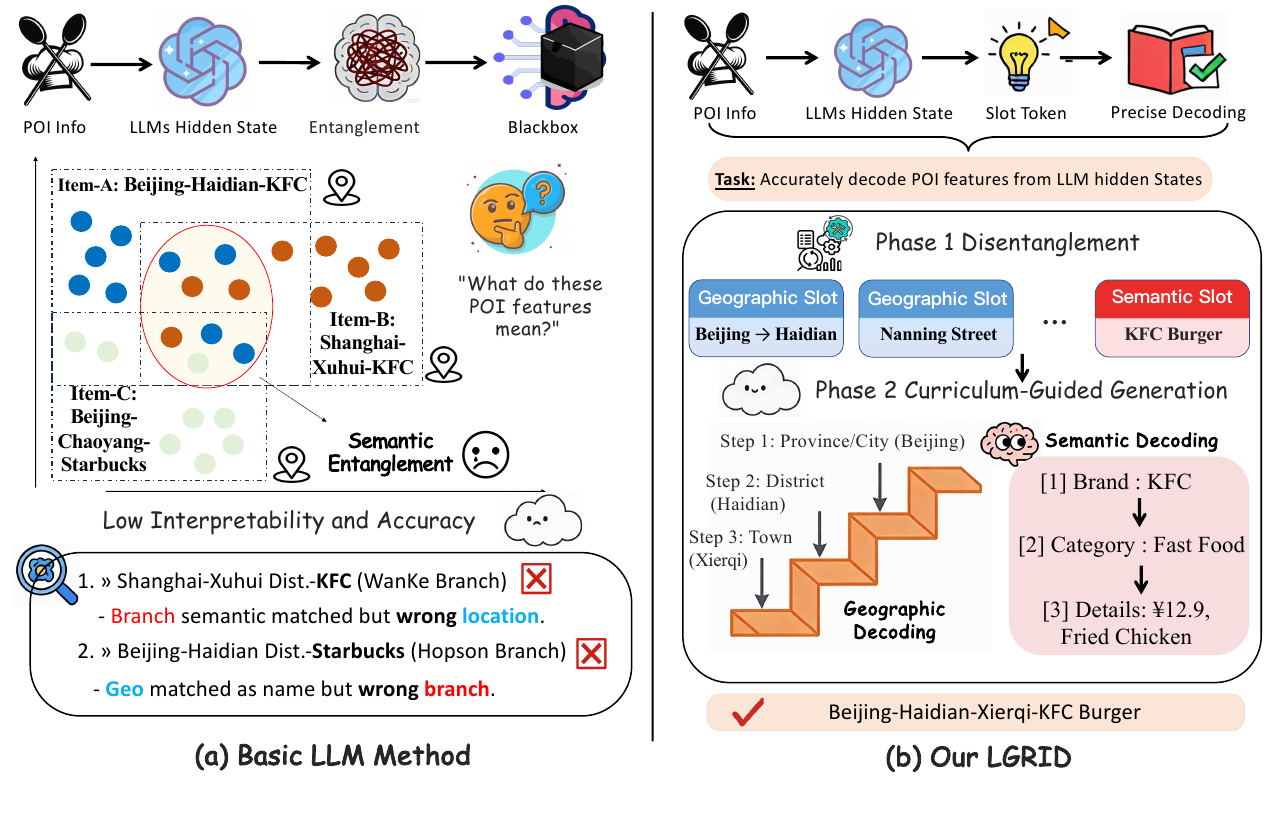}
  \vspace{-1.0cm}
  \caption{Comparison between (a) the single-representation-then-quantization paradigm and (b) our LGRID framework.}
  \label{intro}
  \vspace{-0.4cm}
\end{figure} 

Local-life service recommendation, a representative form of
location-based service (LBS) recommendation, leverages users' location
information and interaction histories to recommend nearby services and
businesses, such as restaurants, food-delivery providers, and local
merchants~\cite{lin2022spatiotemporal,zhao2025spatial}. Major platforms
such as Kuaishou~\cite{jiang2026rest},
Meituan~\cite{lan2025localgpt}, and
Ele.me~\cite{li2023fragment} serve large user populations by matching
them with geographically relevant content and local
services~\cite{jiang2026rest,wang2025fim}. Unlike conventional recommendation
scenarios, LBS recommendation is fundamentally constrained by spatial
reachability~\cite{jiang2025llm,wei2025oneloc,lv2026reasoning}.
Due to offline visiting costs and distance limitations, user decisions typically follow the pattern of \textbf{geographic reachability first and interest matching second}~\cite{jiang2025llm}.

Conventional ID-based recommenders are ill-suited to LBS recommendation because
sparse item IDs encode neither transferable semantics nor explicit
spatial structure~\cite{zhou2019deep}.
Prior studies incorporate spatiotemporal signals into user
interaction sequences, but they typically treat location information as
auxiliary context or encode it as discretized features, rather than
learning item-level representations with explicit spatial
structure~\cite{hu2025dynamic,chen2025next,zhang2025next,
song2025integrating}.
As a result, they struggle to capture spatially correlated user behavior
and the effect of user--item distance on interaction likelihood. 
Since interaction probability typically decreases with distance, spatial
awareness is essential for LBS recommendation~\cite{zhang2025video,
jiang2026rest}.

Recent advances in large language models
(LLMs)~\cite{yang2025qwen3,liu2024deepseek} and Semantic ID (SID)
generation~\cite{rajput2023recommender,zheng2024adapting,
luo2025qarm} offer a promising alternative. LLMs provide transferable semantic priors, while SID methods encode item content or collaborative signals and quantize the resulting representations
into compact token identifiers that replace or augment sparse item IDs.
To improve spatial awareness, recent geo-aware SID methods, including
OneLoc~\cite{wei2025oneloc}, GNPR-SID~\cite{wang2025generative}, and LGSID~\cite{jiang2025llm}, inject or align geographic signals during item
tokenization. 
Nevertheless, these methods typically inject geographic signals into a shared item representation before quantization,
rather than preserving explicit attribute structure~\cite{fu2025forge,
xu2025mmq,zhang2025video,zhang2025enhancing}. Consequently, the quantizer
still operates on a mixed representation of geographic, semantic, and
contextual signals. We refer to this design as the
\textbf{single-representation-then-quantization} paradigm.

As illustrated in Fig.~\ref{intro}, this paradigm introduces two key
bottlenecks in LBS recommendation. First, \textbf{semantic entanglement} at the representation stage degrades SID quality during quantization. When geography, brand, and category are compressed into one latent space, dominant signals such as brand can obscure fine-grained but decision-critical geographic distinctions. 
This is particularly harmful in LBS, where items with similar brand semantics
may correspond to different service regions; for example,
``Beijing--Haidian--KFC'' and ``Shanghai--Xuhui--KFC'' should not receive
overly similar SIDs.
Second, \textbf{black-box} single-representation learning and quantization
limit interpretability: the mixed representation lacks explicit attribute
semantics, making SID positions difficult to attribute to specific geographic
or semantic factors.
This problem is amplified by geographic hierarchy in LBS: province, city, district,
and town define progressively finer service regions, yet existing methods
flatten them into a mixed representation, so the resulting identifiers may
preserve brand semantics while misencoding location or branch information,
as illustrated in Fig.~\ref{intro}(a).

A seemingly straightforward remedy is to encode structured item fields,
such as geography, brand, and category, independently. Although this avoids cross-field mixing, it also blocks the geo-content interactions. Table~\ref{tab:fieldwise_compact} supports this argument: LGRID achieves
$31.4\%$--$88.5\%$ relative MRR gains over independent field-wise
encoding. In LBS, geography not only locates an item but also
contextualizes its semantics; e.g., ``Apple'' in an electronics
district is more likely a retailer than a fruit vendor.
Thus, fused encoding entangles heterogeneous attributes, whereas independent field-wise encoding loses the geo-content interactions. 
SID generation for LBS recommendation therefore requires a framework that
preserves such dependencies through joint LLM encoding and imposes explicit
attribute structure before quantization. Since LLM hidden states have been
shown to encode extractable, decodable, and steerable concepts
~\cite{zou2023representation,xiao2024enhancing,zhao2025steering,
wang2025episodic,song2025towards}, this motivates the central question:
\textit{Can an SID generation framework transform entangled LLM hidden states
into attribute-aligned geographic and semantic slots before quantization?}

To answer this question, we propose \textbf{LGRID}, which formulates SID
generation as \textbf{generative disentanglement before quantization} for
local-life service recommendation. As shown in Fig.~\ref{intro}(b), LGRID follows an encode $\to$ disentangle $\to$ align $\to$ quantize pipeline. It first
encodes each item with an LLM using item descriptions that capture
location-related, merchant, and category cues, thereby preserving
geo-content dependencies. The SD-Block then routes the mixed hidden states into attribute-aligned geographic and semantic slots before quantization. \textbf{Synergistic Alignment Learning} makes these slots LLM-decodable, discriminative, and non-redundant, enabling
attribute-level interpretation and verification. Finally,
\textbf{Dual-Stream Residual Quantization (DSRQ)} separately discretizes
the geographic and semantic slots into compact SIDs with clearer spatial
and semantic structure.

In summary, our major contributions are as follows.
\begin{itemize}[leftmargin=7pt,topsep=0pt]

\item We identify key limitations in the prevailing
\textbf{single-representation-then-quantization} paradigm for SID generation
in local-life service recommendation. Many existing SID methods, including
geo-aware variants, still quantize a shared item representation, leading to
semantic entanglement, black-box, and severe collisions, while
independent field-wise encoding loses essential geo-content dependencies.

\item We propose \textbf{LGRID}, a
\textbf{generative-disentanglement-before-quantization} framework for SID
generation in local-life service recommendation. It instantiates an encode $\to$ disentangle $\to$ align $\to$
quantize pipeline that preserves geo-content dependencies, constructs
attribute-aligned geographic and semantic slots, and separately quantizes
these slots into compact SIDs.

\item Experiments on Kuaishou and Foursquare show that LGRID consistently
improves recommendation performance across diverse recommenders, achieving
up to a $5.44\%$ relative AUC gain. It further predicts coarse geographic
attributes with over $99\%$ accuracy and reduces the full-SID collision rate to 39.9\%.
\end{itemize}

\vspace{-0.3cm}
\section{Related work}
\label{sec:related_work}

\subsection{Item Tokenization}
\label{sec:rw_sid}
Traditional recommendation models represent items using sparse IDs learned
from collaborative signals~\cite{guo2020autodis,wang2017deep,zhou2018deep}.
However, such ID-based representations lack semantic information and rely
heavily on interaction data, making them vulnerable to cold-start and
long-tail scenarios and leading to inefficient memory usage as the ID
vocabulary grows~\cite{rajput2023recommender}. To address these
limitations, recent item tokenization methods leverage LLMs to encode
items into semantic embeddings and quantize them into fixed-length token
sequences, typically following a two-stage paradigm of semantic
representation and quantization.
On the representation side, TIGER~\cite{rajput2023recommender} and
LC-Rec~\cite{zheng2024adapting} extract semantic embeddings from
item-side content, while LETTER~\cite{wang2024learnable}, EAGER~\cite{wang2024eager},
and OneRec~\cite{zhou2025onerec,zhou2025onerec2} further incorporate
collaborative or multimodal signals.
On the quantization side, 
residual-based methods dominate by constructing
coarse-to-fine codes. 
RQ-VAE~\cite{rajput2023recommender} is adopted by
LC-Rec~\cite{zheng2024adapting}, Recbase~\cite{zhou2025recbase},
COBRA~\cite{yang2025sparse}, and GFlowGR~\cite{wang2025gflowgr},
while QARM~\cite{luo2025qarm} and similarity-/n-gram-based variants
\cite{lin2025unified,zheng2025enhancing} further improve codebook
utilization and explore alternative code construction strategies.
Recent studies have incorporated geographic signals into SID generation.
OneLoc~\cite{wei2025oneloc} and GNPR-SID~\cite{wang2025generative}
integrate geographic cues for location-aware recommendation, while
LGSID~\cite{jiang2025llm} aligns LLM representations with spatial signals
through reinforcement learning. Although these methods improve the
geographic awareness of SID representations, geographic signals are still injected into a shared item representation before quantization, rather than used to form explicit attribute-aligned structures. Consequently, the
quantizer operates on an entangled latent vector whose SID positions do not
explicitly correspond to geographic or semantic factors. In contrast, LGRID
constructs attribute-aligned geographic and semantic slots before
quantization and quantizes these slots separately into compact SIDs.

\vspace{-0.3cm}
\subsection{Interpretable LLM Representation}
\label{sec:rw_interp}
LLM-derived item embeddings remain black boxes whose internal
dimensions carry no attribute-level meaning, which prevents the
resulting codes from being inspected or verified. Within
recommendation, LMIndexer~\cite{jin2024language} and
PLUM~\cite{he2025plum} learn semantic IDs directly from LLM
representations, but treat the IDs as flat sequences that encode a
holistic representation, obscuring the boundaries between attribute
levels; more broadly, existing interpretability efforts in LLM-based
recommendation provide holistic, post-hoc explanations for
representations~\cite{li2023personalized,liu2025onerec}, without
exposing the attribute-level structure required by local life
services. In the NLP community, probing
studies~\cite{belinkov2022probing,li2023inference,park2024linear} and
inspection frameworks such as
Patchscopes~\cite{ghandeharioun2024patchscopes} show that LLM hidden
states contain linearly extractable structure, while representation
engineering and activation steering~\cite{zou2023representation,
zhao2025steering,stoehr2024activation,wu2025sharp} further demonstrate
that such structure can be manipulated at inference time. 
However, probing and steering mainly treat interpretability as a post-hoc
or inference-time operation, leaving a gap between what is explained and
what is used for SID construction. LGRID closes this gap by turning
interpretability into a training-time representation-construction objective:
it learns attribute-aligned geographic and semantic slots that are decoded
and aligned during training, and then quantized separately into compact SIDs.

\begin{figure*}[tp]
  \includegraphics[width= \linewidth]{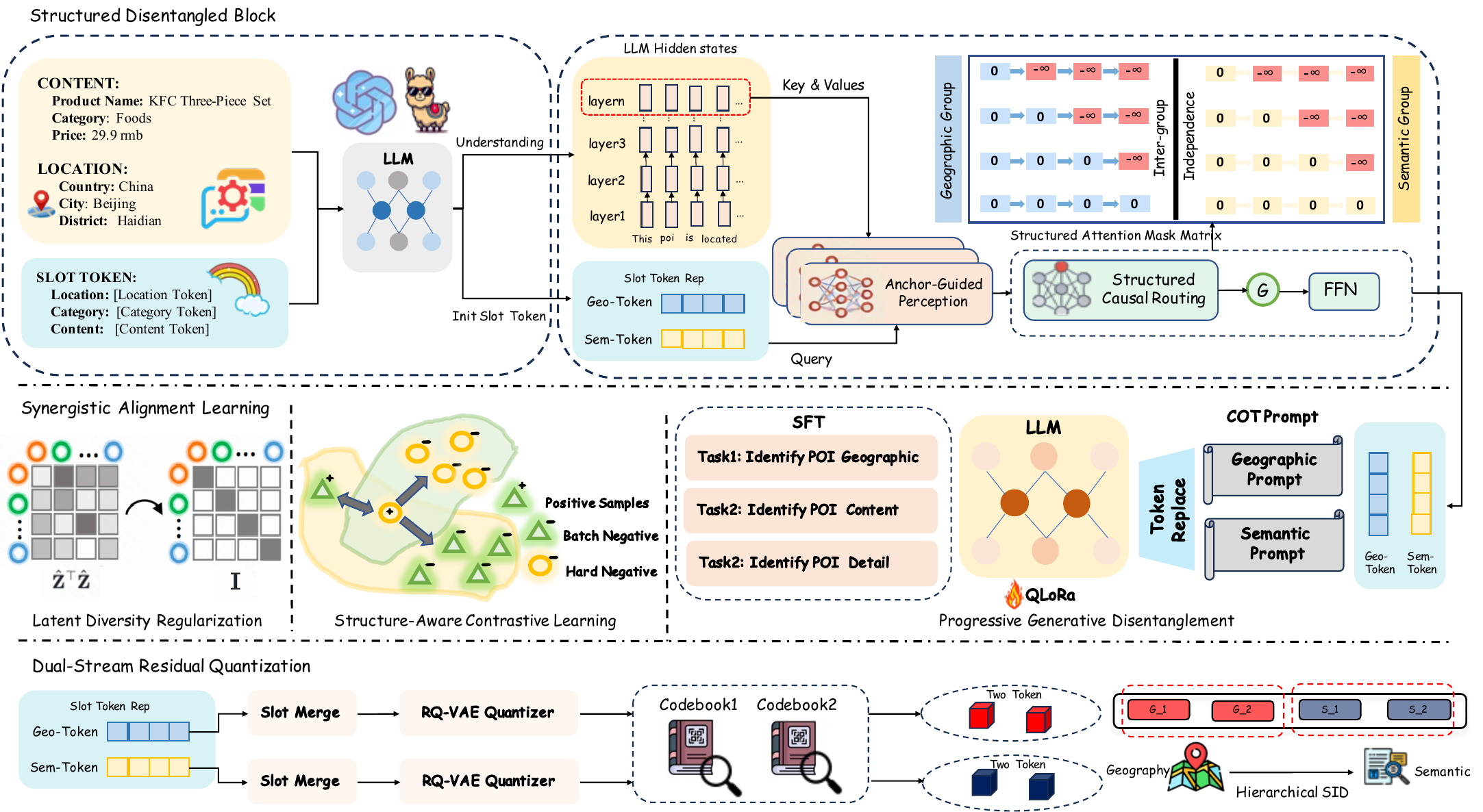}
  \vspace{-0.6cm}
  \caption{Overview of the LGRID framework. The architecture integrates the SD Block to disentangle features, employs Synergistic Alignment Learning to ensure representation quality, and utilizes Dual-Stream RQ to generate compact SIDs.}
  \label{f4}
  \vspace{-0.5cm}
\end{figure*}

\vspace{-0.3cm}
\section{METHODOLOGY}  

\subsection{Problem Definition}
LBS recommendation aims to predict user preferences among candidate items strictly constrained by a spatial radius (e.g., 5 km). Formally, let $\mathcal{U}$ and $\mathcal{V}$ denote the sets of users and items, respectively, with user $u$'s historical sequence represented as $H_u = (v_1, \dots, v_T)$. We define a spatially valid candidate set $\mathcal{V}_u^D \subseteq \mathcal{V}$ comprising only items within a distance limit $D$. Accordingly, the task is formulated as learning a scoring function to estimate the interaction probability for $v \in \mathcal{V}_u^D$ at step $T+1$. LGRID derives interpretable hierarchical item representations that enhance this scoring process.

\vspace{-0.3cm}
\subsection{Overview}
LGRID follows an encode $\to$ disentangle $\to$ align $\to$ quantize
pipeline that targets the failure modes of
single-representation-then-quantization at the representation-construction
stage (Fig.~\ref{f4}). This order is central: instead of quantizing a fused
item representation, LGRID first makes attribute structure explicit and then
discretizes the resulting geographic and semantic slots. An LLM first jointly
encodes POI text into entangled hidden states. Stage~1, the
\textbf{Structured Disentangled Block (SD-Block)}, routes these states into
attribute-aligned geographic and semantic slots. Stage~2,
\textbf{Synergistic Alignment Learning}, makes the slots LLM-decodable,
discriminative, and non-redundant through coarse-to-fine attribute
supervision and contrastive regularization. Finally, Stage~3,
\textbf{Dual-Stream Residual Quantization (DSRQ)}, quantizes the geographic
and semantic slot streams separately into compact Semantic IDs.

\vspace{-0.3cm}
\subsection{Joint LLM Encoding}
Given an item text $\mathcal{S}$ covering location, brand, and category,
the backbone $\mathrm{LLM}_\phi$ produces contextual hidden states
$\mathbf{H} = \mathrm{LLM}_\phi{}   (\mathcal{S}) \in \mathbb{R}^{L\times d}$,
allowing self-attention to capture bidirectional geo-content
dependencies before any decoupling. Although semantically rich,
$\mathbf{H}$ is typically entangled~\cite{hao2024training}, making
naive aggregation (e.g., mean pooling or the \texttt{[CLS]} token)
prone to semantic smoothing and information loss. This entangled
representation motivates the structural decoupling that follows; we
empirically validate joint encoding against independent field-wise
encoding in Section~\ref{sec:exp_fieldwise}.

\vspace{-0.3cm}
\subsection{Structured Disentangled Block}
The SD-Block decouples the entangled hidden states $\mathbf{H}$ into
$N$ attribute-aligned slot tokens through three components: Semantic
Anchor Injection (SAI), Anchor-Guided Perception (AGP), and Structured
Causal Routing (SCR).

\subsubsection{Semantic Anchor Injection}
Unlike standard Q-Formers~\cite{li2023blip} whose randomly initialized
queries have no inherent binding to specific attributes due to the
permutation symmetry of
attention~\cite{locatello2020object,carion2020end,dong2021attention},
we initialize the $N$ query vectors from domain-specific prompt tokens
$\mathcal{T}$ corresponding to attributes (e.g., ``Provincial Unit'',
``Brand Name''), yielding semantic anchors~\cite{lester2021power}:
\vspace{-0.1cm}
\begin{equation}
\label{eq:slot_init}
    \mathbf{Z}^{(0)} = \mathrm{LLM}_\phi(\mathcal{T}) + \mathbf{P}_{\text{pos}},
\vspace{-0.1cm}
\end{equation}
where $\mathbf{P}_{\text{pos}}$ denotes positional embeddings.
Grounding slot tokens in the LLM's semantic space establishes a
one-to-one correspondence between slots and physical attributes. We instantiate $N$ slots: $k$ geographic slots mirroring the administrative hierarchy (province, city, district, town) and $N-k$
semantic slots (e.g., brand and category); the exact attribute list
and anchor strings are given in
Appendix~\ref{app:semantic_anchor_tokens}.

\subsubsection{Anchor-Guided Perception}
AGP employs the slot tokens to probe attribute-specific information
from $\mathbf{H}$. We first inject sinusoidal positional encodings,
$\hat{\mathbf{H}} = \mathbf{H} + \mathbf{P}_{\text{src}}$, then in the
$l$-th layer the slots act as queries in multi-head cross-attention:
\begin{equation}
\hat{\mathbf{A}}^{(l)} = \mathbf{Z}^{(l-1)} +
\mathrm{Dropout}\big(\mathrm{CrossAttn}(\mathbf{Z}^{(l-1)},
\hat{\mathbf{H}}, \hat{\mathbf{H}})\big).
\end{equation}
AGP thus acts as a learnable filter, routing sparse cues (e.g., mentions of ``Chaoyang District'') into the corresponding slots while
suppressing noise. Residual ``soft entanglement'' (e.g., brand names
containing place names) may persist and is resolved by SCR.

\subsubsection{Structured Causal Routing}
\label{sec:scr}
Within each disentanglement layer, the AGP output
$\hat{\mathbf{A}}^{(l)}$ (denoted $\hat{\mathbf{Z}}$ for brevity) is
further refined by Structured Causal Routing (SCR), where ``causal''
refers to the autoregressive attention order within each group.

\noindent\textbf{Structured Attention Mask.}
We partition the $N$ slot tokens into a geographic group
$\mathcal{G}_{\text{loc}} = [0, k)$ and a semantic content group
$\mathcal{G}_{\text{sem}} = [k, N)$; the slot-to-attribute assignment
follows the semantic anchors defined in SAI. To enforce ``intra-group
autoregression and inter-group independence,'' we incorporate a
structured mask $\mathbf{M} \in \mathbb{R}^{N \times N}$ into the
self-attention mechanism:
\vspace{-0.1cm}
\begin{equation}
    \mathbf{M}_{ij} = \begin{cases} 
    0 & \text{if } \exists g \in \{\mathcal{G}_{\text{loc}}, \mathcal{G}_{\text{sem}}\}, \{i, j\} \subseteq g \land j \leq i \\
    -\infty & \text{otherwise.}
    \end{cases}
\vspace{-0.1cm}
\end{equation}
This design imposes two hard constraints: 1) \textbf{Intra-group Autoregression ($j \le i$)} compels tokens to attend only to preceding slots within the same group, aligning the information flow with the
containment order of administrative divisions (e.g., Province $\to$
City);
and 2) \textbf{Inter-group Independence
($\{i, j\} \not\subseteq g$)} blocks cross-group attention paths, preventing feature mixing across groups.

\noindent\textbf{Adaptive Gated Routing.}
We apply Layer Normalization and inject positional encodings
$\mathbf{P}$ to construct the query and key matrices
$\mathbf{Q} = \mathbf{K} = \mathrm{LayerNorm}(\hat{\mathbf{Z}}) + \mathbf{P}$,
with values $\mathbf{V} = \mathrm{LayerNorm}(\hat{\mathbf{Z}})$. The
information flow is governed by the structured mask $\mathbf{M}$:
\vspace{-0.1cm}
\begin{equation}
    \mathbf{S} = \mathrm{Softmax}\left( \frac{\mathbf{Q}\mathbf{K}^\top}{\sqrt{d}} + \mathbf{M} \right) \mathbf{V}.
    \vspace{-0.1cm}
\end{equation}
To avoid gradient instability from hard masking at initialization~\cite{bachlechner2021rezero}, a learnable gate
$\sigma(\alpha)$~\cite{srivastava2015highway} modulates the update:
\vspace{-0.1cm}
\begin{equation}
    \mathbf{Z}_{\text{scr}} = \hat{\mathbf{Z}} + \sigma(\alpha) \cdot \mathrm{Dropout}(\mathbf{S}),
    \vspace{-0.1cm}
\end{equation}
followed by a standard pre-norm FFN block,
$\mathbf{Z}^{(l)} = \mathrm{FFN}(\mathrm{LayerNorm}$ $(\mathbf{Z}_{\text{scr}})) + \mathbf{Z}_{\text{scr}}$,
the final layer's output is denoted $\mathbf{Z}$.

\subsection{Synergistic Alignment Learning}
After the SD-Block produces attribute-aligned slots, Stage~2
(Synergistic Alignment Learning) optimizes them to be LLM-decodable
and discriminative. This stage couples two complementary
objectives -- Progressive Generative Disentanglement (PGD) for
hierarchical semantic order and Structure-Aware Contrastive Learning
(SACL) for fine-grained discriminability---together with a lightweight
diversity regularizer that prevents slot collapse.
\vspace{-0.1cm}
\subsubsection{Progressive Generative Disentanglement} 
PGD makes each slot LLM-decodable through coarse-to-fine attribute supervision. Given only slot representations and an
attribute-specific prompt, the decoder $P_{\text{dec}}$ (instantiated by
the LLM backbone) predicts the attribute value associated with the target
slot, encouraging the slot to be semantically decodable and aligned with
its intended attribute. For each PGD task, the corresponding attribute
label is used only as the supervision target, while the decoder input
contains only the prompt and slot representations. Thus, PGD
serves as a slot-level semantic decoding objective for interpretability,
rather than input-field reconstruction.
PGD follows a curriculum-guided decoding path that respects attribute
dependencies: the geographic chain (Slots~1--4) predicts province $\to$
city $\to$ district $\to$ town, with each finer level conditioned on
coarser geographic context, while the semantic chain (Slots~5--8) is
decoded sequentially (brand $\to$ category $\to$ finer semantic
attributes). To predict the attribute value $y_m$ at the $m$-th level,
we concatenate the slot tokens $\mathbf{Z}_{\leq m}$ with an
attribute-specific decoding prompt $\mathbf{p}_m$, and optimize the
decoder with attribute-level supervision:
\vspace{-0.2cm}
\begin{equation}
    \mathcal{L}_{\text{pgd}}^{(m)} = - \sum_{t=1}^{T_m} \log
    P_{\text{dec}}(y_{m,t} \mid y_{m,<t}, \mathbf{p}_m, \mathbf{Z}_{\leq m}).
\vspace{-0.1cm}
\end{equation}
For instance, task T3 (district level) conditions on the province and city recovered by T1--T2: the input concatenates the instruction
(``\ldots identify the district within the given city\ldots'') with the slot embeddings $\mathbf{Z}_{0:3}$ and an $\langle\mathtt{answer}\rangle$ marker, and the loss is applied only to the target district text, with instruction and slot tokens masked out. By constraining generation to progress from holistic to local, this curriculum-style objective routes attributes into their designated slots without manual annotation; the full task schema and all seven instruction prompts are provided in Appendix~\ref{app:pgd_prompts}.

\vspace{-0.1cm}
\subsubsection{Structure-Aware Contrastive Learning}
While PGD establishes a preliminary hierarchical structure, purely
generative objectives often suffer from discriminative bottlenecks,
leading to over-smoothed representations for structurally similar
entities (e.g., shops in different towns within the same district).
To prevent such over-smoothing, we introduce Structure-Aware
Contrastive Learning (SACL), enforces semantic consistency via positive alignment and injects fine-grained discriminative signals
through structured hard negatives.

\noindent\textbf{Semantic Consistency: Positive Sample Construction.}
To keep slots semantically faithful to the item, we maximize the agreement between the aggregated slot representation
(mean-pooled over slots; anchor view $\mathbf{v}_{\text{anc}}$) and a
reference view derived from ground-truth metadata (target view
$\mathbf{v}_{\text{pos}}$). Concretely, $\mathbf{v}_{\text{pos}}$ is
obtained by feeding a metadata-derived descriptive sentence into the
LLM backbone and aggregating the hidden states via weighted positional
pooling, providing a supervisory signal free from compression loss.

\noindent\textbf{Structured Discrimination: Hierarchical Hard Negative
Mining.}
Standard random sampling often fails to distinguish highly similar
entities. Let $y_{\text{geo}}^{(c)}$, $y_{\text{geo}}^{(f)}$, and
$y_{\text{sem}}$ denote coarse-grained location, fine-grained location,
and semantic content, respectively. For an anchor $x_i$, we construct
three complementary hard-negative sets, each controlling one attribute
while varying another.

\noindent\textbf{(i) Spatial Specificity.}
We select items in the same coarse region but at different fine-grained
locations:
\vspace{-0.1cm}
\begin{equation}
\label{eq:loc}
\mathcal{N}_{\text{loc}}=\{x_j \mid y_{\text{geo}}^{(c)}(x_j)=y_{\text{geo}}^{(c)}(x_i) \land y_{\text{geo}}^{(f)}(x_j)\neq y_{\text{geo}}^{(f)}(x_i)\}.
\end{equation}
\vspace{-0.1cm}
This forces the model to preserve fine-grained spatial distinctions
within the same coarse region.

\noindent\textbf{(ii) Semantic Independence.}
We select items at the same fine-grained location but with different
semantic content:
\vspace{-0.1cm}
\begin{equation}
\label{eq:sem}
\mathcal{N}_{\text{sem}}=\{x_j \mid y_{\text{geo}}^{(f)}(x_j)=y_{\text{geo}}^{(f)}(x_i) \land y_{\text{sem}}(x_j)\neq y_{\text{sem}}(x_i)\}.
\end{equation}
\vspace{-0.1cm}
This prevents geographic context from becoming a shortcut for semantic
discrimination.

\noindent\textbf{(iii) Cross-Region Discrimination.}
We select items with the same semantic content but in different coarse
regions:
\vspace{-0.1cm}
\begin{equation}
\label{eq:ent}
\mathcal{N}_{\text{ent}}=\{x_j \mid y_{\text{sem}}(x_j)=y_{\text{sem}}(x_i) \land y_{\text{geo}}^{(c)}(x_j)\neq y_{\text{geo}}^{(c)}(x_i)\}.
\end{equation}
\vspace{-0.1cm}
This prevents semantically similar items in different regions from
collapsing into indistinguishable representations.

We combine the three sets as
$\mathcal{N}_{\text{hard}}(x_i)=
\mathcal{N}_{\text{loc}}\cup
\mathcal{N}_{\text{sem}}\cup
\mathcal{N}_{\text{ent}}$.
When a target set is empty, we hierarchically relax the corresponding
geographic constraint, e.g., from town to district, to improve coverage
while retaining negative hardness
(Appendix~\ref{app:sacl_details}).

\noindent\textbf{Unified Contrastive Objective.}
We integrate these samples into a unified InfoNCE loss:
\vspace{-0.2cm}
\begin{equation}
    \mathcal{L}_{\text{sacl}} = - \log \frac{E(\mathbf{v}_{\text{pos}})}{E(\mathbf{v}_{\text{pos}}) + \lambda \sum_{\mathbf{v}_n \in \mathcal{N}_{\text{hard}}} E(\mathbf{v}_n) + \sum_{\mathbf{v}_r \in \mathcal{N}_{\text{batch}}} E(\mathbf{v}_r)},
    \vspace{-0.1cm}
\end{equation}
where
$E(\mathbf{v}) = \exp(\mathrm{sim}(\mathbf{v}_{\text{anc}}, \mathbf{v}) / \tau)$
denotes the exponentiated similarity and $\mathcal{N}_{\text{batch}}$
represents in-batch negatives.

\subsubsection{Latent Diversity Regularization.}
To prevent inter-slot redundancy, we impose Latent Diversity
Regularization (LDR), which encourages subspace orthogonality so that
each slot captures non-overlapping features. This is achieved by
minimizing the Frobenius norm of the difference between the Gram
matrix of the $\ell_2$-normalized slot representations
$\tilde{\mathbf{Z}}$ and the identity matrix:
\vspace{-0.1cm}
\begin{equation}
    \mathcal{L}_{\text{ldr}} = \| \tilde{\mathbf{Z}}^\top \tilde{\mathbf{Z}} - \mathbf{I} \|_F^2.
    \vspace{-0.1cm}
\end{equation}

\subsubsection{Total Optimization Objective}
Finally, LGRID is optimized by jointly minimizing these objectives:
\begin{equation}
    \mathcal{L} = \lambda_{1} \sum_{m} \mathcal{L}_{\text{pgd}}^{(m)} + \lambda_{2} \mathcal{L}_{\text{sacl}} + \lambda_{3} \mathcal{L}_{\text{ldr}}.
\end{equation}
\vspace{-0.2cm}

\begin{algorithm}[t]
\caption{LGRID Semantic ID Construction Pipeline}
\label{alg:lgrid_pipeline}
\small 
\begin{algorithmic}[1]
\Require POI text $\mathcal{S}$; backbone $\mathrm{LLM}_\phi$; anchors $\mathcal{T}$; DSRQ codebooks
\Ensure Semantic ID $\mathrm{SID}(v)$
\State \textbf{Encode \& init:} $\mathbf{H}=\mathrm{LLM}_\phi(\mathcal{S})$;\quad $\mathbf{Z}^{(0)}=\mathrm{LLM}_\phi(\mathcal{T})+\mathbf{P}_{\text{pos}}$
\State \textbf{Disentangle:} $\mathbf{Z}=\mathrm{SDBlock}(\mathbf{H},\mathbf{Z}^{(0)})$ \Comment{SAI $\to$ AGP $\to$ SCR}
\State \textbf{Align:} optimize $\mathbf{Z}$ with $\lambda_1\!\sum_m\!\mathcal{L}_{\text{pgd}}^{(m)}+\lambda_2\mathcal{L}_{\text{sacl}}+\lambda_3\mathcal{L}_{\text{ldr}}$
\State \textbf{Quantize:} $\mathbf{C}_{\text{geo}}=\mathrm{RQ}_{\text{geo}}(\mathbf{Z}_{[0:k)})$;\quad $\mathbf{C}_{\text{sem}}=\mathrm{RQ}_{\text{sem}}(\mathbf{Z}_{[k:N)})$
\State \Return $\mathrm{SID}(v)=[c_{\text{geo}}^{(1)},c_{\text{geo}}^{(2)},c_{\text{sem}}^{(1)},c_{\text{sem}}^{(2)}]$
\end{algorithmic}
\end{algorithm}

\vspace{-0.4cm}
\subsection{Dual-Stream Residual Quantization}
Given the verifiable, discriminative slots from Stage~2, DSRQ discretizes them into compact Semantic IDs, carrying the
geographic--semantic separation into the discrete codes.

\vspace{-0.2cm}
\subsubsection{Parallel Dual-Stream Architecture.}
Given the distributional divergence between geographic
($\mathbf{Z}_{\text{geo}}$) and semantic ($\mathbf{Z}_{\text{sem}}$)
features, a shared quantization space often causes codebook
competition. We thus adopt two parallel branches: a geographic branch
mapping $\mathbf{Z}_{\text{geo}}$ to codes $\mathbf{C}_{\text{geo}}$,
and a semantic branch mapping $\mathbf{Z}_{\text{sem}}$ to
$\mathbf{C}_{\text{sem}}$.

\subsubsection{Structured Residual Quantization.}
To ensure high-precision mapping within a limited codebook budget, we
employ Residual Quantization (RQ)~\cite{rajput2023recommender}, which
performs a coarse-to-fine recursive approximation. Each stream's slots
are first aggregated into a single vector $\mathbf{z}$ (by
concatenation followed by projection), which is then decomposed into a
linear combination of $D$ codewords from codebook $\mathcal{C}$ over
$D$ iterations, yielding the reconstruction
$\hat{\mathbf{z}} = \sum_{d=1}^{D} \mathbf{e}_{i_d}$, where
$\mathbf{e}_{i_d} \in \mathcal{C}$. This coarse-to-fine recursion mirrors PGD's hierarchy, preserving granularity in the discrete codes.

DSRQ is trained after Stage~2 converges, with the slot representations
fixed, via reconstruction and commitment losses:
\vspace{-0.1cm}
\begin{equation}
    \mathcal{L}_{\text{RQ}} = \|\mathbf{z} - \hat{\mathbf{z}}\|^2 + \beta \sum_{d=1}^D \|\mathrm{sg}[\mathbf{r}_{d-1}] - \mathbf{e}_{i_d}\|^2,
\vspace{-0.1cm}
\end{equation}
where $\mathrm{sg}[\cdot]$ is the stop-gradient operator and $\mathbf{r}_d = \mathbf{r}_{d-1} - \mathbf{e}_{i_d}$ is the residual at
step $d$ ($\mathbf{r}_0 = \mathbf{z}$). Each stream uses $D=2$ code levels, yielding a 4-token Semantic ID $\mathrm{SID}(v) = [c_{\text{geo}}^{(1)}, c_{\text{geo}}^{(2)},
c_{\text{sem}}^{(1)}, c_{\text{sem}}^{(2)}]$.

\subsubsection{Deployment and Pipeline Summary.}
For non-SID recommendation backbones, the Semantic ID does not replace the original item ID. We embed all four SID codes, sum them into an SID representation, and concatenate it with the original item-ID
embedding:
\vspace{-0.3cm}
\begin{equation}
\mathbf{e}_{v}=\left[\mathbf{e}_{\mathrm{item}}(v);\; \sum_{\ell=1}^{4}\mathbf{e}_{\mathrm{sid}}(c_{v,\ell})\right],
\vspace{-0.1cm}
\end{equation}
where $\mathbf{e}_{\mathrm{item}}(v)$ is the standard item-ID
embedding and $\mathbf{e}_{\mathrm{sid}}(c_{v,\ell})$ is the embedding
of the $\ell$-th SID code. This unified protocol does not require architectural change to the backbone and ensures fair comparison among all recommenders. Algorithm~\ref{alg:lgrid_pipeline} summarizes the full construction pipeline; all steps run offline, and only pre-generated SIDs are used at serving time.

\begin{table*}[t]
\centering
\caption{Performance comparison (AUC) across recommendation backbones on
Kuaishou and Foursquare datasets. The gray rows highlight the proposed
method. ``vs.\ SOTA'' reports the relative improvement over the
best baseline for each backbone.}
\label{tab:main_results}
\vspace{-0.3cm}
\small
\setlength{\tabcolsep}{3.5pt}
\renewcommand{\arraystretch}{1.1}
\definecolor{headergray}{RGB}{245,245,245}
\definecolor{rowgray}{RGB}{240,240,240}
\definecolor{mutedred}{RGB}{176,60,60}

\begin{tabular}{c l *{10}{c}}
\toprule
\rowcolor{headergray}
\textbf{Dataset} & \textbf{Method} & \textbf{GRU4Rec} & \textbf{DIN} & \textbf{SASRec} & \textbf{BERT4Rec} & \textbf{DIEN} & \textbf{SIM} & \textbf{ETA} & \textbf{TWIN} & \textbf{HSTU} & \textbf{RankMixer} \\
\midrule

\multirow{9}{*}{\rotatebox{90}{\textbf{Kuaishou}}}
& Base \cite{yang2025qwen3}
& 0.6324 & 0.5994 & 0.6252 & 0.6408 & 0.6444 & 0.5991 & 0.6036 & 0.6124 & 0.6387 & 0.5995\\
& Res-KMeans \cite{luo2025qarm}
& 0.6377 & 0.6361 & 0.6382 & 0.6418 & 0.6436 & 0.6580 & 0.8259 & 0.6671 & 0.6453 & 0.6048\\
& RQ-VAE \cite{rajput2023recommender}
& 0.6373 & 0.6346 & \underline{0.6529} & \underline{0.6554} & 0.6365 & 0.6468 & \underline{0.8745} & 0.6509 & 0.6448 & 0.6130\\
& RQ-VAE-cos \cite{lin2025unified}
& 0.6454 & 0.6306 & 0.6524 & 0.6536 & 0.6345 & 0.6464 & 0.8612 & 0.6535 & 0.6440 & \underline{0.6131}\\
& RQ-VAE-NGram \cite{zheng2025enhancing}
& 0.6301 & 0.6462 & 0.6465 & 0.6429 & 0.6154 & 0.6596 & 0.7629 & 0.6612 & 0.6336 & 0.6037\\
& LGSID \cite{jiang2025llm}
& \underline{0.6451} & 0.6519 & 0.6334 & 0.6435 & 0.6460 & 0.6576 & 0.8165 & 0.6610 & 0.6433 & 0.6126\\
& LGSID++ \cite{jiang2025llm}
& 0.6436 & \underline{0.6546} & 0.6493 & 0.6538 & \underline{0.6551} & \underline{0.6671} & 0.8327 & \underline{0.6734} & \underline{0.6465} & 0.6019\\

\rowcolor{rowgray}
\cellcolor{white} & \textbf{Ours (LGRID)}
& \textbf{0.6805} & \textbf{0.6604} & \textbf{0.6647} & \textbf{0.6670} & \textbf{0.6617} & \textbf{0.6732} & \textbf{0.8821} & \textbf{0.6818} & \textbf{0.6492} & \textbf{0.6187}\\

& \textit{vs.\ SOTA} ($\Delta\%$)
& {\color{mutedred}\textbf{+5.44\%$\uparrow$}}
& {\color{mutedred}\textbf{+0.89\%$\uparrow$}}
& {\color{mutedred}\textbf{+1.81\%$\uparrow$}}
& {\color{mutedred}\textbf{+1.77\%$\uparrow$}}
& {\color{mutedred}\textbf{+1.01\%$\uparrow$}}
& {\color{mutedred}\textbf{+0.91\%$\uparrow$}}
& {\color{mutedred}\textbf{+0.87\%$\uparrow$}}
& {\color{mutedred}\textbf{+1.25\%$\uparrow$}}
& {\color{mutedred}\textbf{+0.42\%$\uparrow$}}
& {\color{mutedred}\textbf{+0.92\%$\uparrow$}}\\
\midrule

\multirow{9}{*}{\rotatebox{90}{\textbf{Foursquare}}}
& Base \cite{yang2025qwen3}
& 0.7717 & 0.7948 & 0.7712 & \underline{0.7730} & 0.7920 & 0.8198 & 0.8217 & 0.8160 & 0.7746 & 0.7732 \\
& Res-KMeans \cite{luo2025qarm}
& 0.7988 & 0.8406 & 0.7900 & 0.7714 & 0.8570 & 0.8663 & 0.8584 & 0.8662 & \underline{0.7832} & \underline{0.7772} \\
& RQ-VAE \cite{rajput2023recommender}
& 0.7994 & 0.8773 & 0.7882 & 0.7677 & 0.8744 & 0.8927 & 0.8901 & 0.8910 & 0.7736 & 0.7744 \\
& RQ-VAE-cos \cite{lin2025unified}
& 0.7960 & 0.8489 & 0.7883 & 0.7615 & 0.8311 & 0.8707 & 0.8705 & 0.8707 & 0.7739 & 0.7694\\
& RQ-VAE-NGram \cite{zheng2025enhancing}
& 0.7957 & 0.8368 & 0.7914 & 0.7715 & 0.8257 & 0.8683 & 0.8556 & 0.8609 & 0.7773 & 0.7732 \\
& LGSID \cite{jiang2025llm}
& 0.7979 & \underline{0.8781} & 0.7901 & 0.7696 & 0.8746 & 0.8942 & 0.8872 & 0.8921 & 0.7759 & 0.7704 \\
& LGSID++ \cite{jiang2025llm}
& \underline{0.8004} & 0.8728 & \underline{0.7918} & 0.7723 & \underline{0.8772} & \underline{0.8889} & \underline{0.8899} & \underline{0.8913} & 0.7759 & 0.7732 \\

\rowcolor{rowgray}
\cellcolor{white} & \textbf{Ours (LGRID)}
& \textbf{0.8024} & \textbf{0.8785} & \textbf{0.7923} & \textbf{0.7747} & \textbf{0.8910} & \textbf{0.9041} & \textbf{0.9127} & \textbf{0.9107} & \textbf{0.7892} & \textbf{0.7796} \\

& \textit{vs.\ SOTA} ($\Delta\%$)
& {\color{mutedred}\textbf{+0.25\%$\uparrow$}}
& {\color{mutedred}\textbf{+0.05\%$\uparrow$}}
& {\color{mutedred}\textbf{+0.06\%$\uparrow$}}
& {\color{mutedred}\textbf{+0.22\%$\uparrow$}}
& {\color{mutedred}\textbf{+1.57\%$\uparrow$}}
& {\color{mutedred}\textbf{+1.11\%$\uparrow$}}
& {\color{mutedred}\textbf{+2.54\%$\uparrow$}}
& {\color{mutedred}\textbf{+2.08\%$\uparrow$}}
& {\color{mutedred}\textbf{+0.76\%$\uparrow$}}
& {\color{mutedred}\textbf{+0.31\%$\uparrow$}}
\\
\bottomrule
\end{tabular}
\vspace{-0.3cm}
\end{table*}

\vspace{-0.1cm}
\section{Experiment}

\subsection{Experimental Settings}
We evaluate LGRID on a real-world industrial Kuaishou
dataset\footnote{\url{https://www.kuaishou.com/}} and a public LBS benchmark, Foursquare\footnote{\url{https://www.kaggle.com/datasets/rishabhchandra/foursquare-complete-dataset}}.
Both datasets contain geographic and textual item information; for
Foursquare, we derive coarse location descriptions from latitude and
longitude. Detailed dataset statistics are provided in
Appendix~\ref{app:dataset}.

We compare LGRID with representative recommendation backbones, including
DIN~\cite{zhou2018deep}, DIEN~\cite{zhou2019deep},
ETA~\cite{chen2021end}, SIM~\cite{pi2020search},
TWIN~\cite{si2024twin}, HSTU~\cite{zhai2024actions},
RankMixer~\cite{zhu2025rankmixer}, GRU4Rec~\cite{hidasi2015session},
BERT4Rec~\cite{sun2019bert4rec}, and
SASRec~\cite{kang2018self}. We also compare with recent SID
methods, including RQVAE~\cite{rajput2023recommender},
Res-KMeans~\cite{luo2025qarm}, RQVAE-cos~\cite{lin2025unified},
RQVAE-NGram~\cite{zheng2025enhancing}, and
LGSID/LGSID++~\cite{jiang2025llm}. 
LGSID++ uses G-GRPO~\cite{shao2024deepseekmath} instead of
G-DPO~\cite{rafailov2023direct} to improve spatial awareness.

We employ Qwen3-8B~\cite{qwen3technicalreport} as the LLM backbone and
fine-tune it with LoRA~\cite{hu2022lora} on all linear layers. The
SD-Block compresses hidden states into 8 structured slot tokens. LGRID
is integrated into each recommendation backbone by fusing the generated
semantic IDs with original sparse ID embeddings. For recommendation
training, we set the batch size to 10,240 and the embedding dimension of
users, items, and semantic IDs to 8; remaining hyperparameters and
implementation details are provided in Appendix~\ref{app:implementation}.
Training is optimized by AdamW~\cite{loshchilov2017decoupled}.

\newcommand{\imp}[1]{{\cellcolor{lightgray} \textbf{#1}}}

\begin{table}[t]
\centering
\small
\caption{Full-corpus POI self-retrieval. We report MRR, Recall (Rec), and Hit at $K=\{1, 10, 50\}$.}
\vspace{-0.3cm}
\label{tab:recall_res}
\renewcommand{\arraystretch}{1.05}
\setlength{\tabcolsep}{1.35pt}
\resizebox{\columnwidth}{!}{%
\begin{tabular}{ll c cc cc cc}
\toprule
\multirow{2}{*}{\textbf{Dim.}} & \multirow{2}{*}{\textbf{Model}} & \multirow{2}{*}{\textbf{MRR}} &
\multicolumn{2}{c}{\textbf{@1}} & \multicolumn{2}{c}{\textbf{@10}} & \multicolumn{2}{c}{\textbf{@50}} \\
\cmidrule(lr){4-5} \cmidrule(lr){6-7} \cmidrule(lr){8-9}
& & & Rec & Hit & Rec & Hit & Rec & Hit \\
\midrule
\multirow{3}{*}{\textbf{City}}
& Base & 0.9150 & 0.892 & 0.892 & 0.745 & 0.956 & 0.543 & 0.979 \\
& \cellcolor{lightgray}\textbf{LGRID} & \imp{0.9443} & \imp{0.925} & \imp{0.925} & \imp{0.877} & \imp{0.977} & \imp{0.819} & \imp{0.991} \\
& {\textbf{Gain}}
& {\color{mutedred}\textbf{+3.2\%$\uparrow$}}
& {\color{mutedred}\textbf{+3.7\%$\uparrow$}}
& {\color{mutedred}\textbf{+3.7\%$\uparrow$}}
& {\color{mutedred}\textbf{+17.7\%$\uparrow$}}
& {\color{mutedred}\textbf{+2.2\%$\uparrow$}}
& {\color{mutedred}\textbf{+50.8\%$\uparrow$}}
& {\color{mutedred}\textbf{+1.2\%$\uparrow$}} \\
\midrule
\multirow{3}{*}{\textbf{Dist.}}
& Base & 0.7402 & 0.689 & 0.689 & 0.450 & 0.834 & 0.249 & 0.891 \\
& \cellcolor{lightgray}\textbf{LGRID} & \imp{0.8832} & \imp{0.842} & \imp{0.842} & \imp{0.765} & \imp{0.952} & \imp{0.679} & \imp{0.976} \\
& {\textbf{Gain}}
& {\color{mutedred}\textbf{+19.3\%$\uparrow$}}
& {\color{mutedred}\textbf{+22.2\%$\uparrow$}}
& {\color{mutedred}\textbf{+22.2\%$\uparrow$}}
& {\color{mutedred}\textbf{+70.0\%$\uparrow$}}
& {\color{mutedred}\textbf{+14.1\%$\uparrow$}}
& {\color{mutedred}\textbf{+172.7\%$\uparrow$}}
& {\color{mutedred}\textbf{+9.5\%$\uparrow$}} \\
\midrule
\multirow{3}{*}{\textbf{Brand}}
& Base & 0.9532 & 0.936 & 0.936 & 0.923 & 0.978 & 0.902 & 0.994 \\
& \cellcolor{lightgray}\textbf{LGRID} & \imp{0.9713} & \imp{0.958} & \imp{0.958} & \imp{0.939} & \imp{0.993} & \imp{0.921} & \imp{0.997} \\
& {\textbf{Gain}}
& {\color{mutedred}\textbf{+1.9\%$\uparrow$}}
& {\color{mutedred}\textbf{+2.4\%$\uparrow$}}
& {\color{mutedred}\textbf{+2.4\%$\uparrow$}}
& {\color{mutedred}\textbf{+1.7\%$\uparrow$}}
& {\color{mutedred}\textbf{+1.5\%$\uparrow$}}
& {\color{mutedred}\textbf{+2.1\%$\uparrow$}}
& {\color{mutedred}\textbf{+0.3\%$\uparrow$}} \\
\bottomrule
\end{tabular}
}
\vspace{-0.2cm}
\end{table}

\vspace{-0.2cm}
\subsection{Overall Performance}
Table~\ref{tab:main_results} reports AUC results on the Kuaishou and
Foursquare datasets. LGRID achieves the best performance on both datasets
and across all recommendation backbones, indicating that the generated SIDs
are compatible with diverse sequential and industrial ranking models. On
Kuaishou, LGRID outperforms the strongest SID baseline by up to $5.44\%$.
Similar gains are observed on Foursquare, with clear improvements on ETA
and TWIN ($+2.54\%$ and $+2.08\%$). These results suggest that the benefit
of LGRID is not tied to a specific recommender architecture; instead, its
attribute-aligned SIDs provide reusable geographic and semantic signals that
complement downstream ranking models.

\begin{table}[t]
\centering
\caption{Attribute decoding accuracy (\%) across datasets.}
\label{tab:dim-acc}
\vspace{-0.3cm}
\small
\setlength{\tabcolsep}{3.5pt}
\renewcommand{\arraystretch}{1.05}
\begin{tabular}{lccccccc}
\toprule
\textbf{Dataset} & \textbf{Country} & \textbf{Prov.} & \textbf{City} &
\textbf{Dist.} & \textbf{Town} & \textbf{Brand} & \textbf{Cat.} \\
\midrule
\textbf{Kuaishou}   & --    & 99.02 & 99.66 & 99.12 & 92.34 & 86.62 & 97.00 \\
\textbf{Foursquare} & 99.97 & 99.93 & --    & 95.03 & --    & --    & 96.77 \\
\bottomrule
\end{tabular}
\vspace{-0.1cm}
\end{table}

\begin{table}[t]
\centering
\small
\caption{Joint vs.\ field-wise structured encoding (MRR).}
\label{tab:fieldwise_compact}
\vspace{-0.3cm}
\setlength{\tabcolsep}{3.6pt}
\renewcommand{\arraystretch}{1.03}
\begin{tabular}{lcccccc}
\toprule
\textbf{Method} & \textbf{Prov.} & \textbf{City} & \textbf{Dist.} &
\textbf{Town} & \textbf{Brand} & \textbf{Cat.} \\
\midrule
Field-wise & 0.751 & 0.719 & 0.633 & 0.334 & 0.204 & 0.052 \\
\cellcolor{lightgray}\textbf{LGRID} & \imp{0.987} & \imp{0.990} & \imp{0.922} & \imp{0.479} & \imp{0.328} & \imp{0.098} \\
\textbf{Gain}
& {\color{mutedred}\textbf{+31.4\%$\uparrow$}}
& {\color{mutedred}\textbf{+37.7\%$\uparrow$}}
& {\color{mutedred}\textbf{+45.7\%$\uparrow$}}
& {\color{mutedred}\textbf{+43.3\%$\uparrow$}}
& {\color{mutedred}\textbf{+60.9\%$\uparrow$}}
& {\color{mutedred}\textbf{+88.5\%$\uparrow$}} \\
\bottomrule
\end{tabular}
\vspace{-0.3cm}
\end{table}

\vspace{-0.3cm}
\subsection{Representation Quality and Interpretability}
\noindent\textbf{Attribute-consistent POI self-retrieval.}
Table~\ref{tab:recall_res} evaluates POI embedding quality by retrieving
top-k ranked POIs from the full corpus and measuring attribute consistency
with the query. LGRID improves both semantic and geospatial consistency:
Brand MRR increases from $0.9532$ to $0.9713$, and Brand Recall@50
improves from $0.902$ to $0.921$. For geographic attributes, the gains
become larger as granularity becomes finer: City-level MRR improves by
$3.2\%$, while District-level MRR improves by $19.3\%$ and
District-level Recall@50 improves by $172.7\%$. These results indicate
that LGRID preserves hierarchical spatial dependencies while maintaining
semantic coherence. They further support the design choice of disentangling
after joint LLM encoding: the model preserves geo-content interactions while
avoiding their collapse into a single fused representation before
quantization. Full results are reported in
Appendix~\ref{app:full_retrieval}.

\begin{figure}[h] 
    \centering
    \includegraphics[width=1.0\linewidth]{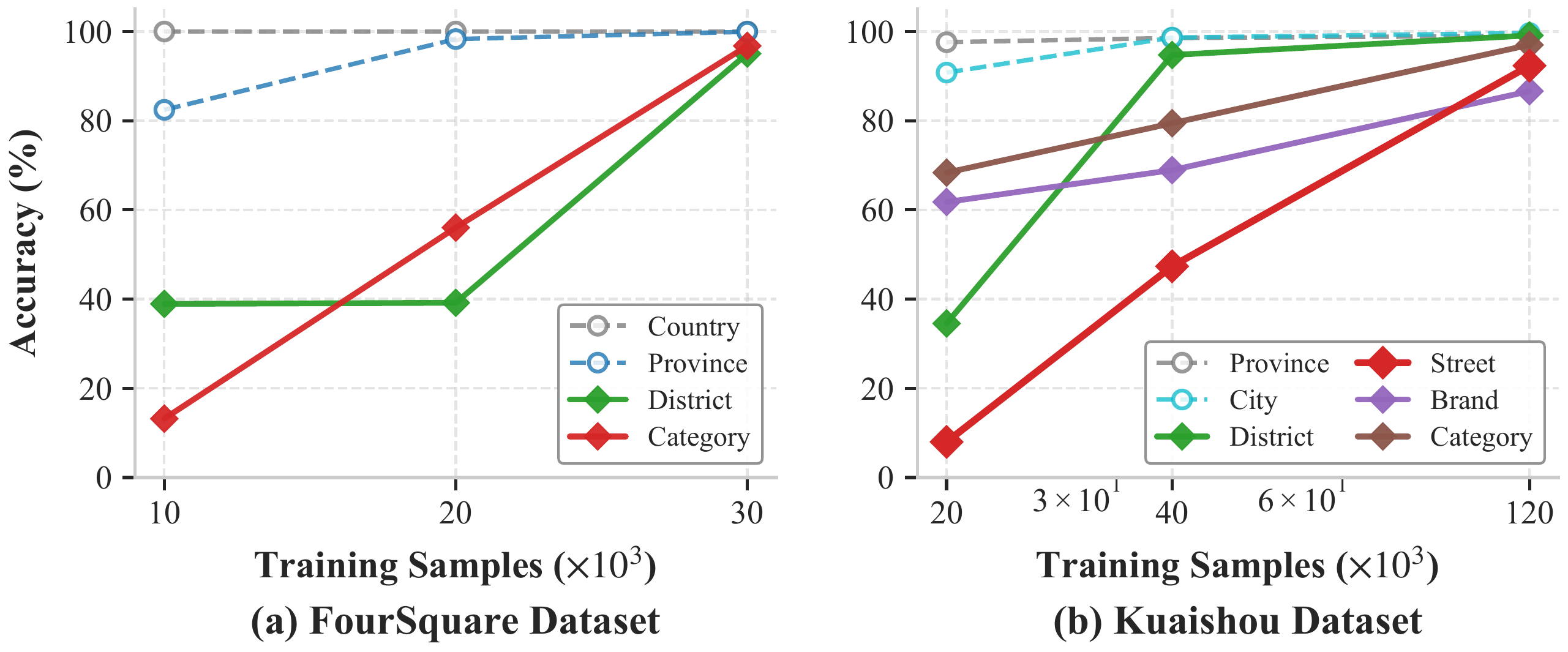}
    \vspace{-0.8cm}
    \caption{\textbf{Scaling Law Analysis.} Performance trajectories on (a) Foursquare and (b) Kuaishou datasets.} 
    \label{fig:scaling_law}
    \vspace{-0.3cm}
\end{figure}

\noindent\textbf{Attribute interpretability.}
Table~\ref{tab:dim-acc} evaluates slot interpretability through attribute decoding. Given the learned slot embeddings and a task prompt, the decoder predicts the corresponding attribute label.
LGRID recovers available coarse spatial attributes with over $99\%$ accuracy on both datasets and
remains strong on fine-grained and semantic attributes in Kuaishou,
including Town ($92.34\%$) and Brand ($86.62\%$).
These results suggest that SD-Block preserves hierarchical semantics
while SCR reduces cross-domain interference, allowing the learned slots
to retain physically interpretable attribute information rather than
acting as opaque latent codes.

\noindent\textbf{Joint encoding necessity.}
\label{sec:exp_fieldwise}
Table~\ref{tab:fieldwise_compact} compares LGRID with a field-wise
structured encoding baseline that separately encodes POI geographic,
brand, and category fields; the full protocol and metrics are provided in
Appendix~\ref{app:structured_field_baseline}. LGRID improves all six
attributes, and the gains increase with geographic granularity, from
$+31.4\%$ on Province to $+45.7\%$ on District.
Semantic attributes benefit even more, with $+60.9\%$ on Brand and $+88.5\%$ on Category. These results suggest that
separately encoding structured fields cuts off geo--content interactions
needed for fine-grained disambiguation, while joint encoding followed by
structural decoupling better preserves both spatial hierarchy and
semantic context.

\vspace{0.05cm}
\noindent\textbf{SID collision analysis.}
\label{sec:collision}
Table~\ref{tab:collision_utilization} reports full-SID collision rate and
cumulative $n$-gram code utilization on Kuaishou; formal definitions are
given in Appendix~\ref{app:codebook_collision}. Existing SID baselines
suffer from severe full-SID collisions ($98.0\%$ for RQ-VAE and $97.0\%$
for LGSID). Although LGSID nearly saturates first-level utilization
($U_1=99.0$), its deeper utilization collapses ($U_{1:2}=4.0$,
$U_{1:3}=0.05$), indicating that later code levels provide limited
additional discrimination. In contrast, LGRID reduces the collision rate to
$39.9\%$ and improves $U_{1:2}$ and $U_{1:3}$ to $100.0$ and $77.3$,
respectively. This contrast supports disentanglement before quantization:
high first-level utilization alone cannot prevent full-SID collisions when
subsequent code levels are learned from the same fused representation.

\begin{table}[h]
\centering
\small
\caption{SID utilization and full-SID collision rate (\%).}
\vspace{-0.3cm}
\label{tab:collision_utilization}
\setlength{\tabcolsep}{6pt}
\begin{tabular}{lcccc}
\toprule
\textbf{Method} & \textbf{$U_1\uparrow$} & \textbf{$U_{1:2}\uparrow$} &
\textbf{$U_{1:3}\uparrow$} & \textbf{Collision$\downarrow$} \\
\midrule
RQ-VAE & 54.4 & 1.9 & 0.05 & 98.0 \\
LGSID & 99.0 & 4.0 & 0.05 & 97.0 \\
\cellcolor{lightgray}\textbf{LGRID} & \imp{100.0} & \imp{100.0} & \imp{77.3} & \imp{39.9} \\
\bottomrule
\end{tabular}
\vspace{-0.2cm}
\end{table}

\noindent\textbf{Scaling law analysis.}
Fig.~\ref{fig:scaling_law} examines how attribute interpretability scales
with training data. LGRID shows a clear coarse-to-fine pattern:
coarse-grained attributes such as Province and City
converge quickly, exceeding $97\%$ accuracy with only $20\mathrm{k}$
Kuaishou samples. 
In contrast, fine-grained Town remains low under small-scale training ($<40\mathrm{k}$ samples, about $7.98\%$), but rises sharply after sufficient data coverage and reaches $92.34\%$ with $120\mathrm{k}$ curated POI-text training samples. This trend suggests that fine-grained spatial disentanglement is data-sensitive, while LGRID's hierarchy supports coarse-to-fine slot learning.

\vspace{-0.2cm}
\subsection{Ablation Study}

\begin{figure}[t]
  \centering
  \includegraphics[width=0.9\linewidth]{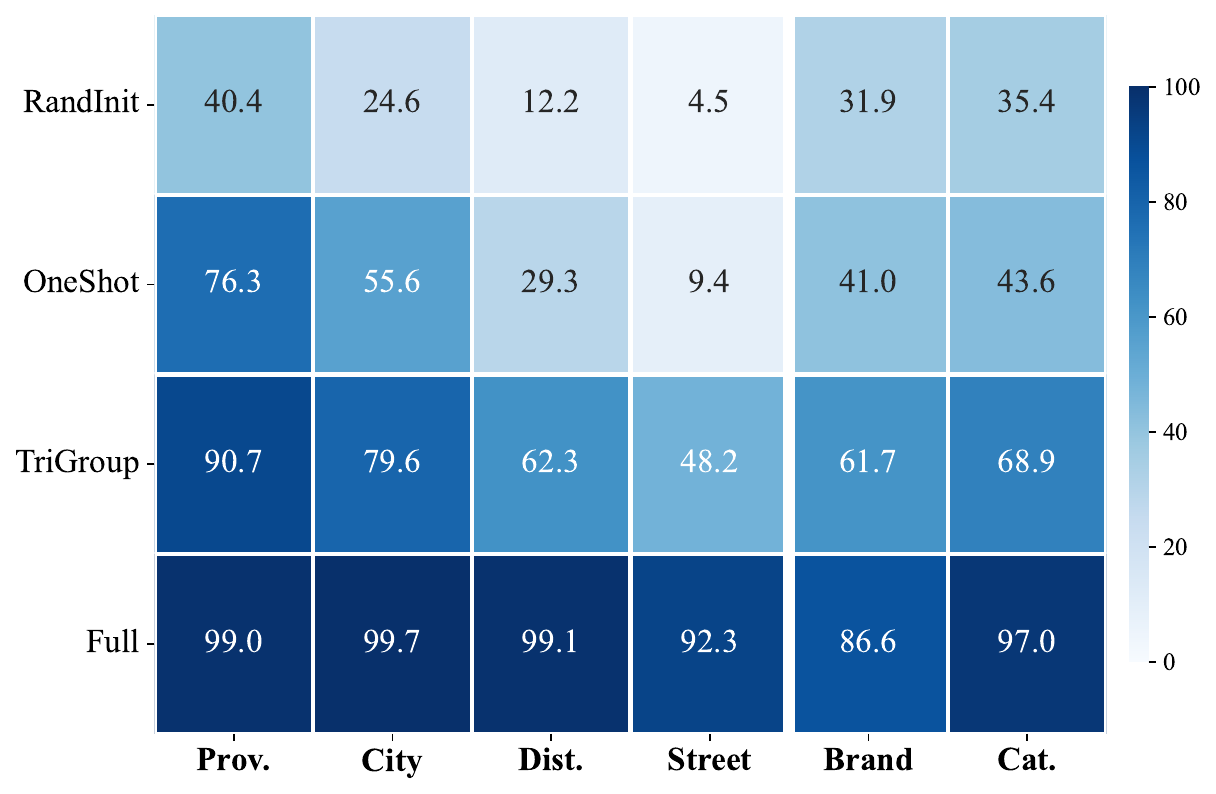}
  \vspace{-0.4cm}
  \caption{Ablation study of learning strategies. The separator distinguishes spatial and semantic attributes.}
  \label{fig:ablation_heatmap}
  \vspace{-0.3cm}
\end{figure}

\noindent\textbf{Learning-strategy ablation.}
Fig.~\ref{fig:ablation_heatmap} ablates semantic anchors and decoding
strategies. Random slot initialization collapses on fine-grained
\textit{Town} recovery ($4.51\%$), showing that SAI stabilizes
token--attribute alignment. One-shot decoding is also weak
($9.36\%$), while coarse domain grouping improves \textit{Town} to
$48.17\%$ but remains far below Full LGRID. This indicates that grouping
tokens is insufficient without the coarse-to-fine dependency modeled by
PGD, which uses coarse slots as contextual prompts and reaches
$92.34\%$ on \textit{Town}. Additional robustness results are in
Appendix~\ref{app:prompt_robustness}.

\noindent\textbf{Design-level ablation.}
Table~\ref{tab:ablation_merge} evaluates SD-Block aggregation and SACL
sampling designs. In the SD-Block group, linear projection fails on all
attributes, indicating that simple projection cannot organize raw LLM hidden
states into structured slots. Q-Former and Cross-Att improve over this
baseline but remain weak on fine-grained spatial and semantic attributes;
Full LGRID further improves \textit{Town} and \textit{Brand} accuracy over
Cross-Att by $78.21$ and $33.89$ points, respectively, suggesting that
Structured Causal Routing better preserves hierarchical spatial semantics.
In the SACL group, vanilla sampling yields limited fine-grained
discrimination (\textit{Town}: $37.17\%$), whereas hierarchical hard
negatives raise it to $83.51\%$. Full LGRID further improves
\textit{Town}, \textit{Brand}, and \textit{Category} by $8.83$, $7.67$,
and $7.57$ points over HierHard, showing that fallback sampling complements
hard-negative mining.

\begin{table}[h]
\centering
\caption{Design-level ablation on attribute recovery accuracy.}
\label{tab:ablation_merge}
\vspace{-0.25cm}
\small
\setlength{\tabcolsep}{2.8pt}
\renewcommand{\arraystretch}{1.05}
\begin{tabular}{@{}llrrrrrr@{}}
\toprule
\textbf{Ablation} & \textbf{Variant} &
\textbf{Prov.} & \textbf{City} & \textbf{Dist.} &
\textbf{Town} & \textbf{Brand} & \textbf{Cat.} \\
\midrule
\multicolumn{2}{@{}l}{\cellcolor{lightgray}\textbf{Full LGRID}} &
\imp{99.02} & \imp{99.66} & \imp{99.12} &
\imp{92.34} & \imp{86.62} & \imp{97.00} \\
\midrule
\multirow{3}{*}{\textbf{SD-Block}}
& Linear     & 0.00  & 0.00  & 0.00  & 0.00  & 0.00  & 0.00  \\
& Q-Former   & 75.14 & 42.37 & 25.47 & 6.85  & 16.65 & 24.42 \\
& Cross-Att  & 87.56 & 63.62 & 44.58 & 14.13 & 52.73 & 59.87 \\
\midrule
\multirow{2}{*}{\textbf{SACL}}
& Simple     & 89.36 & 79.42 & 56.53 & 37.17 & 42.66 & 51.87 \\
& HierHard   & 94.52 & 91.35 & 88.22 & 83.51 & 78.95 & 89.43 \\
\bottomrule
\end{tabular}
\vspace{-0.1cm}
\end{table}

\begin{figure*}[t]  
  \centering
  \includegraphics[width=0.95\textwidth]{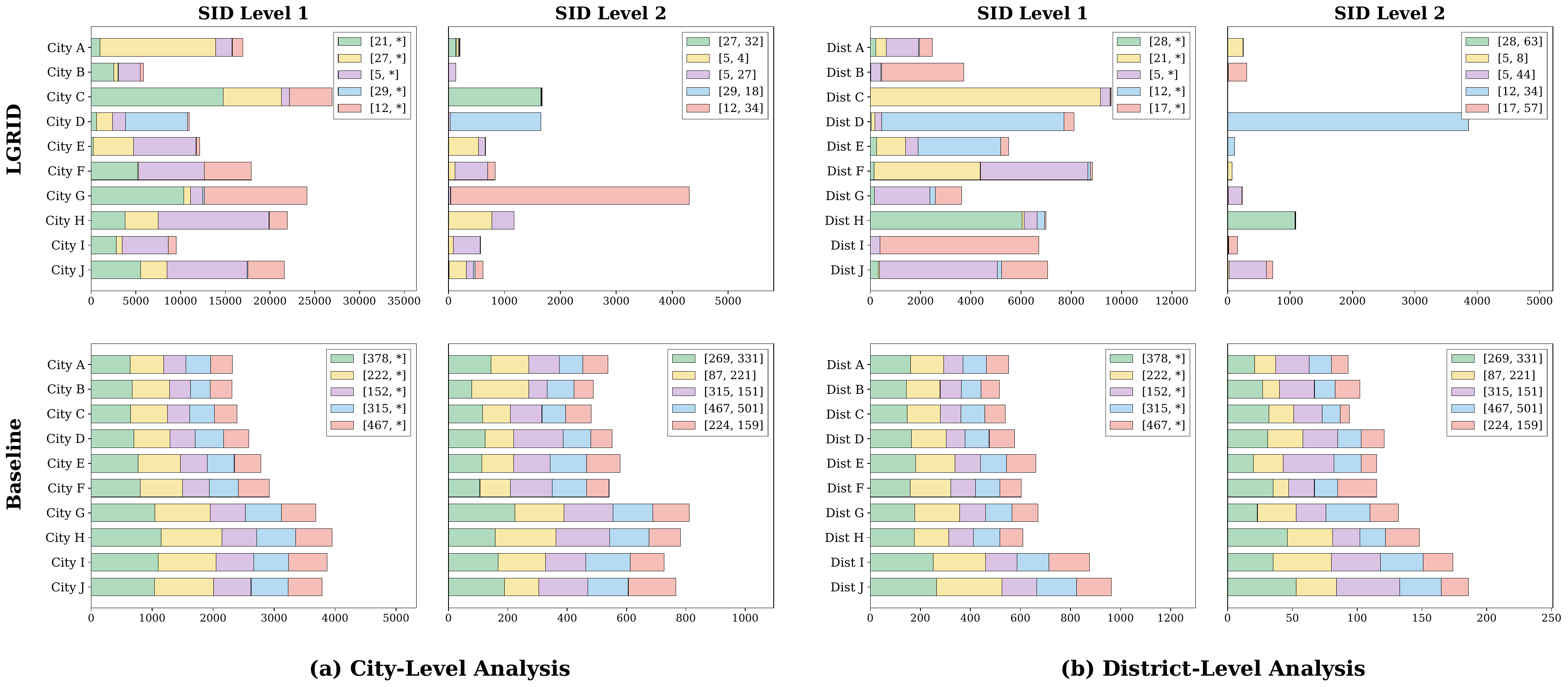}
  \vspace{-0.4cm} 
  \caption{Hierarchical city and district frequency distribution of LGRID for different SID prefixes}
  \label{fig:bar}
  \vspace{-0.4cm}
\end{figure*}

\noindent\textbf{Component-level ablation.}
Table~\ref{tab:failure_mode_ablation} shows that LGRID components play
complementary roles. Replacing SCR with the Q-Former-style fully connected
interaction in Table~\ref{tab:ablation_merge} causes the largest
degradation, reducing \textit{Prov.} Acc and \textit{Brand} Acc by
$23.88$ and $69.97$ percentage points, and \textit{Dist.} R@50 by
$0.388$. Removing $\mathcal{L}_{\mathrm{PGD}}$,
$\mathcal{L}_{\mathrm{SACL}}$, and $\mathcal{L}_{\mathrm{LDR}}$ mainly
affects slot decodability, semantic discrimination/retrieval, and
inter-slot redundancy, respectively. These results show that LGRID is not
a redundant module stack: SCR creates structured slots, PGD assigns
verifiable attribute semantics, SACL improves fine-grained discrimination,
and LDR discourages redundancy, together preparing the slots for
dual-stream quantization.

\begin{table}[h]
\centering
\caption{Component-level ablation on Kuaishou. Acc is in \%; R@50 denotes attribute-consistent retrieval.}
\label{tab:failure_mode_ablation}
\vspace{-0.3cm}
\small
\setlength{\tabcolsep}{6pt}
\renewcommand{\arraystretch}{1.05}
\begin{tabular}{@{}lccc@{}}
\toprule
\textbf{Variant} & \textbf{Prov.\ Acc} & \textbf{Brand Acc} & \textbf{Dist.\ R@50} \\
\midrule
\cellcolor{lightgray}\textbf{Full LGRID} & \imp{99.02} & \imp{86.62} & \imp{0.679} \\ 
w/o SCR & 75.14 & 16.65 & 0.291 \\ 
w/o $\mathcal{L}_{\mathrm{pgd}}$  & 78.40 & 44.80 & 0.452 \\
w/o $\mathcal{L}_{\mathrm{sacl}}$ & 93.80 & 48.70 & 0.336 \\
w/o $\mathcal{L}_{\mathrm{ldr}}$  & 98.30 & 77.80 & 0.598 \\
\bottomrule
\end{tabular}
\vspace{-0.2cm}
\end{table}

\vspace{-0.2cm}
\subsection{Efficiency and Case Study}
\label{sec:efficiency_case}
\noindent\textbf{Efficiency.}
LGRID's extra cost is confined to offline SID construction
($21.80$ vs.\ $13.97$ ms/item, $1.56\times$ over RQ-VAE). Since SIDs are
pre-generated and only looked up during serving, online input preparation
remains nearly unchanged ($0.50$ vs.\ $0.48$ ms/request, $1.04\times$).
Detailed timing results are reported in Appendix~\ref{app:efficiency}.

\begin{figure}[t]
  \centering
  \includegraphics[width=\linewidth]{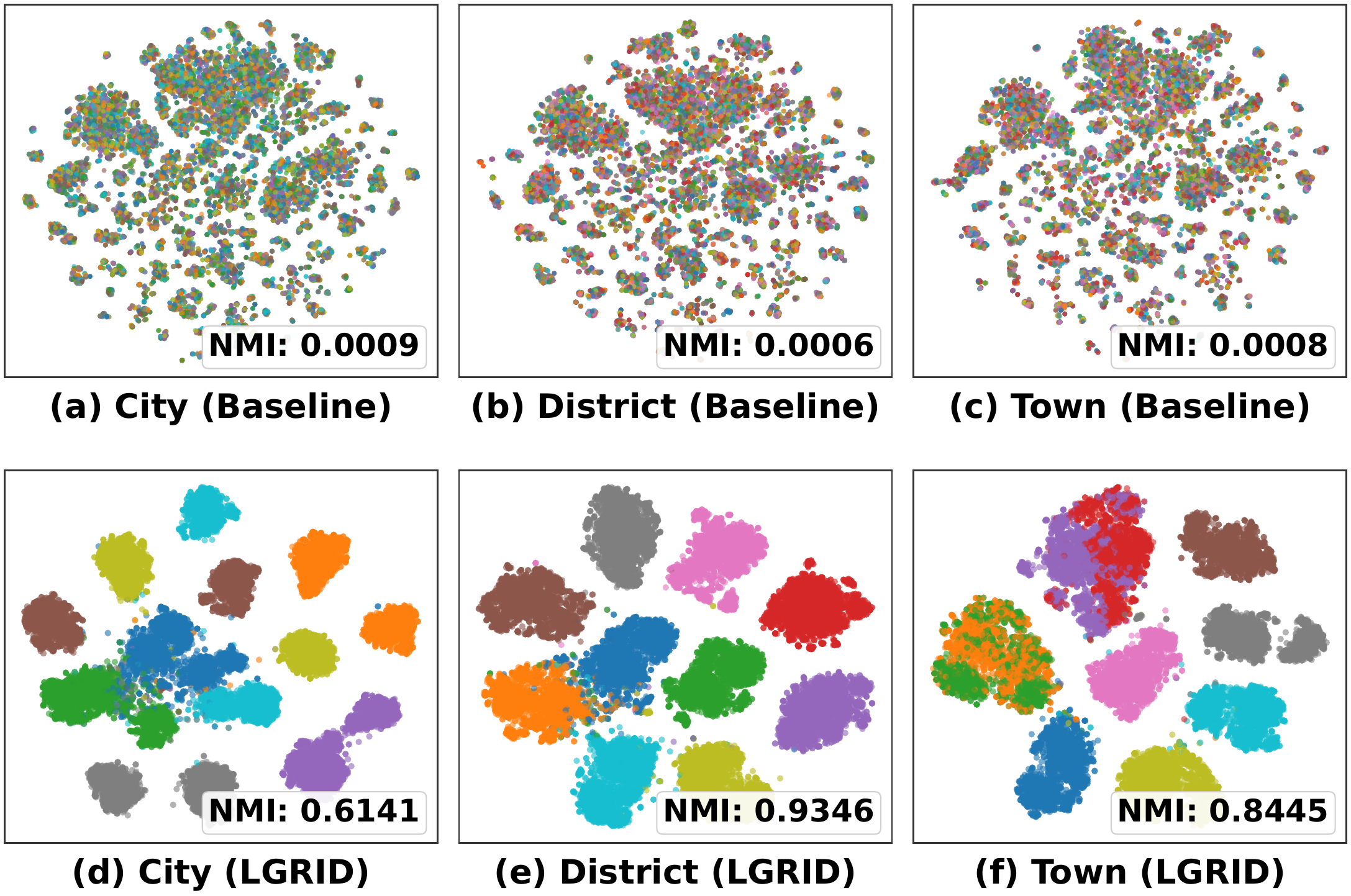}
  \vspace{-0.75cm} 
  \caption{T-SNE visualization of items around cluster centroids across different methods.}
  \label{fig:tsne}
  \vspace{-0.3cm}
\end{figure}

\noindent\textbf{Case study.}
Fig.~\ref{fig:bar} shows top-frequency SID prefixes across cities and
districts. LGRID exhibits a coarse-to-fine spatial hierarchy:
first-level prefixes capture coarse city-level regularities, whereas
second-level prefixes further concentrate on finer district-level
regions. In contrast, the baseline shows high prefix utilization across spatial
units, but these prefixes provide weak spatial discrimination, making its
SIDs more susceptible to code collapse and collisions. Fig.~\ref{fig:tsne} provides a
complementary embedding-space view. Baseline embeddings mix samples from
different geographic units and yield near-zero District NMI ($0.0006$),
whereas LGRID forms compact and well-separated geographic clusters,
reaching a District NMI of $0.9346$. Additional brand/category prefix
visualizations are provided in
Appendix~\ref{app:additional_sid_visualization}. Together, these results
suggest that LGRID maintains online efficiency while learning more discriminative SIDs with hierarchical semantics.

\vspace{-0.2cm}
\section{Conclusions}
In this work, we reframed SID generation for local-life service
recommendation around representation construction before quantization.
Existing methods often fuse geographic hierarchy, brand and category semantics,
and textual context into one representation before discretization. LGRID
instead introduces a \textbf{Generative Disentanglement}
framework, instantiated as an encode $\to$ disentangle $\to$ align $\to$
quantize pipeline. Joint LLM encoding preserves geo-content dependencies;
SD-Block and Synergistic Alignment Learning construct decodable,
discriminative, and non-redundant geographic and semantic slots; and DSRQ
separately quantizes these slots into compact SIDs. Experiments on Kuaishou
and Foursquare show consistent gains, including up to a $5.44\%$ relative
AUC improvement, over $99\%$ attribute-decoding accuracy for coarse-grained
geographic fields, and a much lower full-SID collision rate than LGSID
($39.9\%$ vs. $97.0\%$), supporting explicit attribute structure before
quantization.

\bibliographystyle{unsrt}
\bibliography{Main.bib}

\clearpage
\begin{appendices}


\section{Reproducibility Details}
\label{app:reproducibility}
All industrial data are anonymized and used only for offline evaluation;
user identifiers and exact personal trajectories are not released. This
appendix provides implementation and evaluation details for LGRID. We
first describe the datasets, hyperparameters, evaluation protocol, and
efficiency measurement. We then provide additional method specifications,
including semantic anchor initialization, PGD prompts, SCR masks, SACL
sampling, and dual-stream residual quantization. Finally, we report
extended ablations, retrieval analyses, collision/utilization metrics,
prompt robustness results, and additional SID-prefix visualizations.

\subsection{Dataset Statistics}
\label{app:dataset}

Table~\ref{tab:dataset-stats} summarizes the datasets used in our
experiments. Kuaishou is a large-scale industrial dataset with 10M
user--item interactions, 6.67M users, 2.25M items, and 1,000 categories.
It contains rich POI metadata, including geographic attributes and textual
descriptions. Foursquare is a public LBS dataset with 244K interactions,
1,084 users, 38K items, and 401 categories. Since Foursquare does not
provide the same level of explicit POI text as the industrial corpus, we
derive coarse location descriptions from latitude and longitude. These two
datasets allow us to evaluate LGRID under both large-scale industrial and
sparse public LBS settings.

\begin{table}[h]
  \centering
  \small
  \caption{Dataset statistics.}
  \vspace{-0.4cm}
  \label{tab:dataset-stats}
  \setlength{\tabcolsep}{6pt}
  \renewcommand{\arraystretch}{1.05}
  \begin{tabular}{lrrrr}
    \toprule
    \textbf{Dataset} & \textbf{Interactions} & \textbf{Users} & \textbf{Items} & \textbf{Categories} \\
    \midrule
    Kuaishou   & 10,000,000 & 6,673,350 & 2,251,418 & 1,000 \\
    Foursquare & 244,896    & 1,084     & 38,334    & 401   \\
    \bottomrule
  \end{tabular}
  \vspace{-0.1cm}
\end{table}

\vspace{-0.1cm}
\subsection{Implementation and Hyperparameters}
\label{app:implementation}

\textbf{Curated POI-text subset for SID construction.}
The industrial Kuaishou corpus contains large-scale user--item
interactions and rich POI metadata. For LGRID SID-construction
finetuning, we curate a POI-text subset with $120{,}000$ training POIs
and $5{,}000$ held-out validation POIs. The subset is balanced across
categories and excludes POIs with missing, erroneous, or semantically
incomplete descriptions. Each POI is represented by source item text and
allowed descriptive item information. Standardized geographic labels
(\textit{Province}, \textit{City}, \textit{District}, and \textit{Town})
and semantic labels (\textit{Brand} and \textit{Category}) serve as PGD
supervision targets and keys for constructing structured contrastive
pairs. For a PGD task targeting attribute $y_r$, the textual form of
$y_r$ is excluded from the task input and appears only as supervised
output. We mine $K=6$ structured hard negatives per anchor POI.

\textbf{LGRID SID-construction fine-tuning.}
We fine-tune Qwen3-8B with LoRA applied to all linear projection layers,
using rank $16$, scaling factor $32$, and dropout $0.05$. The SD-Block
contains two cross-attention layers followed by one causal self-attention
layer, each with $16$ heads, and compresses LLM hidden states into $8$
structured slot tokens.
Training follows a two-stage schedule. Stage 1 trains slot--attribute
alignment and attribute decoding for $6$ epochs, with learning rates
$8.0\times10^{-5}$ for LoRA modules and $1.6\times10^{-4}$ for the
compressor. Stage 2 trains fine-grained discrimination for $4$ epochs
with learning rate $5.0\times10^{-5}$, limiting drift from the pretrained
representation.
For SID construction, we use per-device batch size $12$ with $4$
gradient-accumulation steps on $4$ GPUs, giving an effective batch size of
$192$. This setting is used only for POI-text-based SID construction and
is separate from downstream recommendation training. We optimize the model
with AdamW and a cosine learning-rate scheduler with warmup ratio $0.05$.
The attribute-decoding loss weights are $1.6$ for coarse-grained
attributes and $1.8$ for fine-grained attributes. 
The contrastive loss weight is linearly warmed up to $0.66$, and the
orthogonality/diversity regularization weight is $0.05$. For InfoNCE, the
initial temperature is $0.07$ and is clipped to $[0.05,0.20]$ during
training. All SID-construction experiments are implemented with PyTorch
and HuggingFace Transformers and run on four NVIDIA L20 GPUs with 48GB
memory. Table~\ref{tab:hyperparameters} summarizes the hyperparameters
used for LGRID.

\begin{table}[t]
\centering
\small
\caption{Hyperparameters for LGRID.}
\vspace{-0.3cm}
\label{tab:hyperparameters}
\renewcommand{\arraystretch}{1.08}
\setlength{\tabcolsep}{3pt}

\begin{tabular}{
@{}
>{\raggedright\arraybackslash}p{0.19\linewidth}
>{\raggedright\arraybackslash}p{0.45\linewidth}
>{\raggedright\arraybackslash}p{0.3\linewidth}
@{}
}
\toprule
\textbf{Category} & \textbf{Hyperparameter} & \textbf{Value} \\
\midrule

\multirow{4}{*}{\makecell[l]{Model\\config}}
& Backbone & Qwen3-8B \\
& LoRA rank / scale & $r=16$, $\alpha=32$ \\
& LoRA dropout & $0.05$ \\
& LoRA target modules & All linear layers \\
\midrule

\multirow{2}{*}{SD-Block}
& Attention heads & 16 \\
& Slot tokens & 8 \\
\midrule

\multirow{5}{*}{Optimization}
& Optimizer & AdamW \\
& Scheduler & Cosine \\
& Warmup ratio & $0.05$ \\
& Per-device batch size & 12 \\
& Effective global batch size & 192 \\
\midrule

\multirow{3}{*}{Stage 1}
& Objective focus & Alignment \\
& Epochs & 6 \\
& LR (compressor / LoRA) & $1.6{\times}10^{-4}$ / $8.0{\times}10^{-5}$ \\
\midrule

\multirow{3}{*}{Stage 2}
& Objective focus & Fine-grained tuning \\
& Epochs & 4 \\
& LR (global) & $5.0{\times}10^{-5}$ \\
\midrule

\multirow{5}{*}{\makecell[l]{Loss\\weights}}
& Coarse attribute-decoding  & $1.6$ \\
& Fine-grained attribute-decoding & $1.8$ \\
& Contrastive loss & Warmup $\to 0.66$ \\
& Orthogonality / diversity & $0.05$ \\
& InfoNCE temperature &  $[0.05,0.20]$ \\
\midrule

\multirow{2}{*}{Hardware}
& GPUs & 4 GPUs \\
& GPU memory & 48GB each \\
\bottomrule
\end{tabular}
\vspace{-0.1cm}
\end{table}

\begin{table}[h]
\centering
\small
\caption{Efficiency comparison between RQ-VAE and LGRID. Offline construction is measured in ms/item.}
\vspace{-0.3cm}
\label{tab:efficiency_comparison}
\setlength{\tabcolsep}{6pt}
\begin{tabular}{lccc}
\toprule
\textbf{Stage} & \textbf{RQ-VAE} & \textbf{LGRID} & \textbf{Ratio} \\
\midrule
Offline SID construction & 13.97 & 21.80 & $1.56\times$ \\
Online input preparation & 0.48 & 0.50 & $1.04\times$ \\
\bottomrule
\end{tabular}
\vspace{-0.2cm}
\end{table}

\textbf{Downstream recommendation training.}
After SIDs are generated, we integrate them into recommendation backbones
by fusing SID embeddings with original sparse ID embeddings. This
stage uses the generated SIDs as features and is separate from
SID-construction finetuning. For downstream recommendation training, we
use batch size $10{,}240$ and set the embedding dimension of users, items,
and SIDs to $8$. Each prediction tower uses an MLP with dimensions
$[32,16,1]$. We use AdamW with learning rate $0.1$ and a StepLR scheduler
that decays it by $0.9$ every $500$ steps.

\vspace{-0.1cm}
\subsection{Candidate-Set and Evaluation Protocol}
\label{app:evaluation_protocol}

We evaluate LGRID under two recommendation settings. For both datasets, all methods use the same train/validation/test split,
candidate set, and evaluation labels. SID construction is performed
offline before recommendation training, and the downstream recommender
uses only generated SID tokens and standard item-ID embeddings. We report
AUC for recommendation and attribute-consistent retrieval metrics for
representation analysis; no interaction labels are used during SID
construction.

\textbf{Kuaishou spatially constrained ranking.}
Following the LGSID evaluation protocol, we construct the candidate set
with a fixed offline distance threshold $D=15$ km. This threshold is used
only for offline benchmarking and is not a modeling assumption of LGRID.
In production, distance filtering is scenario-dependent and may vary by
category, brand, user context, and serving policy. All methods are
evaluated on the same candidate set.

\textbf{Foursquare full ranking.}
Foursquare lacks a production candidate generator and dynamic distance
policy. We therefore follow the standard full-ranking protocol used in
sequential recommendation studies. Geographic information is still
captured through check-in locality and location-derived POI features.

\vspace{-0.1cm}
\subsection{Efficiency Measurement}
\label{app:efficiency}
Table~\ref{tab:efficiency_comparison} compares the runtime of RQ-VAE and
LGRID. Offline SID construction is reported in milliseconds per item, and
online input preparation in milliseconds per request. LGRID adds moderate
offline overhead ($1.56\times$ over RQ-VAE), while online preparation is
nearly unchanged ($1.04\times$), since generated SIDs are pre-computed and
served by lookup.

\vspace{-0.1cm}
\section{Additional Method Details}
\label{app:method_details}

\subsection{Semantic Anchor Tokens}
\label{app:semantic_anchor_tokens}

In Eq.~\eqref{eq:slot_init}, the initial slot tokens
$\mathbf{Z}^{(0)}$ are obtained by encoding short domain-specific anchor
phrases with the frozen LLM. These phrases are not additional supervision
labels and do not contain target answers. They provide semantically
meaningful initial directions for the slots and reduce permutation
ambiguity during slot learning. Table~\ref{tab:semantic_anchor_tokens}
lists the anchors used in our implementation.

\begin{table}[t]
\centering
\small
\caption{Semantic anchor phrases used for slot initialization.}
\vspace{-0.3cm}
\label{tab:semantic_anchor_tokens}
\setlength{\tabcolsep}{4pt}
\renewcommand{\arraystretch}{1.05}
\begin{tabular}{cll}
\toprule
\textbf{Slot} & \textbf{Anchor} & \textbf{Role} \\
\midrule
\multicolumn{3}{l}{\textit{Geographic stream}} \\
0 & Province unit & Province / state \\
1 & City unit & City / prefecture \\
2 & District location & District / county \\
3 & Town & Town / township / street \\
\midrule
\multicolumn{3}{l}{\textit{Semantic stream}} \\
4 & Brand name & Brand identity \\
5 & Primary category & Business category \\
6 & Hot-selling products & Products / services \\
7 & Business characteristics & Products, services, and style \\
\bottomrule
\end{tabular}
\vspace{-0.2cm}
\end{table}

The industrial POI metadata is in Chinese, so the implementation uses
Chinese anchor phrases with the same meanings. For readability,
Table~\ref{tab:semantic_anchor_tokens} reports English translations. In
all experiments, each anchor phrase is fed to the backbone tokenizer and
its final-token embedding is used to initialize the corresponding slot.

\vspace{-0.2cm}
\subsection{Relationship to Geo-aware SID Methods}
\label{app:lgsid_difference}
Recent geo-aware SID methods, such as LGSID, improve a single item
representation by incorporating geographic signals before hierarchical
quantization. LGRID differs in the representation paradigm before
quantization. It first encodes item descriptions with an LLM to preserve geo-content dependencies, and then structurally decouples the resulting hidden states
into interpretable geographic and semantic slots. Quantization is finally applied to separated streams rather than to a
single mixed vector.

\begin{table*}[t]
\centering
\small
\caption{Paradigm-level comparison between geo-aware SID generation and LGRID.}
\vspace{-0.35cm}
\label{tab:lgsid_lgrid_difference}
\setlength{\tabcolsep}{4pt}
\renewcommand{\arraystretch}{1.18}
\begin{tabular}{
@{}
>{\RaggedRight\arraybackslash}m{0.15\textwidth}
>{\RaggedRight\arraybackslash}m{0.39\textwidth}
>{\RaggedRight\arraybackslash}m{0.40\textwidth}
@{}}
\toprule
\textbf{Aspect} &
\textbf{Geo-aware SID methods, e.g., LGSID} &
\textbf{LGRID} \\
\midrule
Representation &
A single geo-enhanced item embedding. &
Multiple LLM-decodable slots with explicit geographic and semantic roles. \\

Geo modeling &
Coordinate- and distance-aware alignment in one latent space. &
Coarse-to-fine geographic hierarchy encoded as slots 0--3. \\

Semantic modeling &
Geography and content remain mixed before quantization. &
Geography and content interact during LLM encoding and are then structurally decoupled. \\

Interpretability &
SID codes are effective but largely black-box. &
Slots can be decoded into province, city, district, town, brand, and category. \\

Quantization &
Hierarchical quantization over one representation stream. &
Dual-stream residual quantization over separated geographic and semantic streams. \\

Limitation addressed &
Improves geographic consistency of SIDs. &
Reduces geo--semantic entanglement and code collision. \\
\bottomrule
\end{tabular}
\vspace{-0.2cm}
\end{table*}

Therefore, LGRID should not be viewed as adding extra modules to a
geo-aware SID baseline. Its main difference is that disentanglement is
performed before SID construction, so the final codes are generated from
structured, LLM-decodable slots rather than from an opaque mixed vector.

\vspace{-0.3cm}
\subsection{PGD Prompts and SFT Tasks}
\label{app:pgd_prompts}
Progressive Generative Disentanglement (PGD) is implemented as seven
conditional attribute-decoding tasks over learned slot embeddings. Each
task follows an instruction--slot--answer format: an instruction prefix,
a task-specific slot slice, an \texttt{<answer>} marker, and the target
attribute text used only as the supervised output. The loss is applied
only to the target text; instruction tokens, conditioning context, and
continuous slot embeddings are masked out. PGD therefore makes slot
semantics decodable and verifiable, rather than reconstructing an input
field.
For geographic tasks, PGD follows a coarse-to-fine decoding protocol.
During evaluation, higher-level context is generated from preceding PGD
predictions rather than oracle labels. Specifically, T2 conditions on the
province predicted by T1, T3 conditions on the province and city predicted
by T1--T2, and T4 conditions on the province, city, and district predicted
by T1--T3. The current target attribute is used only to compute decoding
accuracy. Table~\ref{tab:pgd_sft_tasks} summarizes the task schema, and
Tables~\ref{tab:pgd_prompts_geo} and~\ref{tab:pgd_prompts_entity} list
the prompts used in our implementation.

\begin{table*}[t]
\centering
\small
\caption{PGD task schema. The slot range specifies accessible slot embeddings, and the target field serves only as supervised output.}
\label{tab:pgd_sft_tasks}
\vspace{-0.2cm}
\setlength{\tabcolsep}{4pt}
\renewcommand{\arraystretch}{1.05}
\begin{tabular*}{\textwidth}{@{\extracolsep{\fill}}lllll@{}}
\toprule
\textbf{Task} & \textbf{Family} & \textbf{Slot input} &
\textbf{Supervision target} & \textbf{Conditioning principle} \\
\midrule
T1 & Geo--Province & $\mathbf{Z}_{0:1}$ & Province / state & Decode the coarsest geographic attribute. \\
T2 & Geo--City & $\mathbf{Z}_{0:2}$ & City / prefecture & Decode city with province-level context. \\
T3 & Geo--District & $\mathbf{Z}_{0:3}$ & District / county & Decode district with province and city context. \\
T4 & Geo--Town & $\mathbf{Z}_{0:4}$ & Town / township / street & Decode micro-location with all coarser geo context. \\
T5 & Entity--Brand & $\mathbf{Z}_{4:5}$ & Brand & Decode standardized merchant or POI brand. \\
T6 & Entity--Category & $\mathbf{Z}_{5:6}$ & Category & Decode the most specific available business category. \\
T7 & Entity--Detail & $\mathbf{Z}_{6:8}$ & Detail text & Decode products, services, and business characteristics. \\
\bottomrule
\end{tabular*}
\vspace{-0.2cm}
\end{table*}

For task $r$, the training pair is:
\vspace{-0.2cm}
\begin{equation}
\mathcal{I}_{r}
= \mathrm{Template}(r,\mathcal{C}_{r})
\oplus \mathbf{Z}_{\mathcal{S}_{r}}
\oplus \langle\mathtt{answer}\rangle,
\quad
\mathcal{O}_{r}=y_{r},
\vspace{-0.2cm}
\end{equation}
where $\mathcal{C}_{r}$ denotes the conditioning context,
$\mathbf{Z}_{\mathcal{S}_{r}}$ is the task-specific slot slice, and
$y_r$ is the ground-truth attribute text. The loss is applied only to
$\mathcal{O}_r$.

\noindent
\textbf{Concrete example.}
The district-level task T3 illustrates how the instruction, conditioning
context, slot evidence, and loss mask are instantiated. During training, the higher-level geographic context in T2--T4 is used
for curriculum supervision. During evaluation and all reported PGD
decoding accuracies, this context is generated autoregressively from
earlier PGD predictions, so no oracle label at the current or finer level
is provided.

\begin{center}
\setlength{\fboxsep}{6pt}
\colorbox{gray!10}{%
\begin{minipage}{0.90\linewidth}
\small
\textbf{Example: District-level PGD training pair (T3).}
\begin{itemize}[leftmargin=*]
    \setlength{\itemsep}{2pt}
    \setlength{\parsep}{0pt}
    \setlength{\parskip}{0pt}

\item \textbf{Metadata record for supervision:} the POI is associated
with province $\{p\}$, city $\{c\}$, district $\{d\}$, town $\{t\}$,
brand $\{b\}$, and category $\{g\}$. These fields define supervision
targets and curriculum contexts, not free-form strings copied into the
slot input.

\item \textbf{Conditioning context}
$\mathcal{C}_{\mathrm{T3}}$: province $\{p\}$ and city $\{c\}$.

\item \textbf{Training pair:}
\begin{itemize}[leftmargin=*]
    \setlength{\itemsep}{0pt}
    \item \textit{Input:} instruction text $\oplus$
    $\mathcal{C}_{\mathrm{T3}}$ $\oplus$
    $\mathbf{Z}_{0:3}$ $\oplus$ \texttt{<answer>}
    \item \textit{Supervision target:} district text $\{d\}$
    \item \textit{Loss mask:} applied only to $\{d\}$; instruction,
    context, and continuous slot embeddings are masked out.
\end{itemize}
\end{itemize}
\end{minipage}}
\end{center}

The geographic tasks follow a cumulative coarse-to-fine order: task T$m$
receives the slot prefix up to the target level and the corresponding
higher-level geographic context. The semantic tasks use task-specific slot
slices because brand, category, and detail attributes do not form a strict
administrative containment hierarchy. T7 uses $\mathbf{Z}_{6:8}$ to
decode the two-slot detail field.

\vspace{-0.15cm}
\subsection{Structured Causal Routing Mask}
\label{app:scr_mask_details}

SCR uses a grouped causal mask over eight slots. Slots 0--3 form the
geographic stream, and slots 4--7 form the semantic stream. Within each
stream, a slot can attend to itself and earlier slots; cross-stream
attention is blocked to reduce leakage between geographic and semantic
factors:
\vspace{-0.2cm}
\begin{equation}
\label{eq:scr_mask}
\mathbf{M}_{ij}=
\begin{cases}
0, & \text{if } i,j\in[0,4) \text{ and } j\le i,\\
0, & \text{if } i,j\in[4,8) \text{ and } j\le i,\\
-\infty, & \text{otherwise.}
\end{cases}
\end{equation}
For eight slots, the visibility matrix forms two independent lower-triangular blocks, enforcing causal routing within and isolation between the geographic and semantic streams. The former follows a coarse-to-fine hierarchy, while the latter follows the brand--category--detail schema.

\subsection{Structure-Aware Contrastive Learning}
\label{app:sacl_details}

For each anchor POI, SACL constructs one positive view and multiple
structured hard negatives. The positive view is generated from the POI
metadata as a structured core sentence and encoded by the frozen LLM. This metadata-derived sentence is used only to construct the contrastive target view during SID-construction training; it is not used as an input feature in downstream recommendation.
Hard negatives are sampled from inverted indices built over province, city,
district, town, and brand fields. We use three types of hard negatives:

\begin{itemize}

\item \textbf{Spatial specificity negatives}: same district but different town; if unavailable, same city but different district.
\item \textbf{Semantic independence negatives}: same town but different brand; if unavailable, same city or same province but different brand.
\item \textbf{Entity invariance negatives}: same brand but different district; if unavailable, same brand but different city or province.
\end{itemize}
When a hard-negative pool is too small, we fill the remaining slots with
random negatives. This fallback stabilizes training for sparse long-tail
POIs while preserving structured hardness when available.

\vspace{-0.1cm}
\subsection{Dual-stream Residual Quantization}
\label{app:dsrq_details}

After structural decoupling, LGRID obtains a geographic representation
$\mathbf{Z}_{\mathrm{geo}}$ and a semantic representation
$\mathbf{Z}_{\mathrm{sem}}$. Because these representations have
different semantics and distributional properties, using a shared
quantization space may cause codebook competition or semantic confusion.
DSRQ therefore uses two parallel quantization branches: a geographic
branch maps $\mathbf{Z}_{\mathrm{geo}}$ to geographic codes
$\mathbf{C}^{g}$, and a semantic branch maps $\mathbf{Z}_{\mathrm{sem}}$
to semantic codes $\mathbf{C}^{s}$.

Each branch uses residual quantization. Given an input vector
$\mathbf{z}$, residual quantization approximates it through a
coarse-to-fine sequence of codewords:
\vspace{-0.2cm}
\begin{equation}
\hat{\mathbf{z}}=\sum_{d=1}^{D}\mathbf{e}_{k_d},
\quad
\mathbf{e}_{k_d}\in\mathcal{C}_d,
\vspace{-0.3cm}
\end{equation}
where $\mathcal{C}_d$ is the codebook at residual level $d$. At each
level, the selected codeword approximates the current residual, and the
remaining residual is passed to the next level. The final full SID is
formed by serializing the geographic and semantic code sequences.

\begin{table*}[t]
\centering
\small
\vspace{-0.3cm}
\caption{Geographic stream PGD instruction prompts (T1--T4). Each prompt is fed as the instruction prefix.}
\label{tab:pgd_prompts_geo}
\vspace{-0.3cm}

\setlength{\tabcolsep}{4pt}
\begin{tabular}{p{0.06\textwidth} p{0.88\textwidth}}
\toprule
\textbf{Task} & \textbf{Instruction prompt} \\
\midrule
T1 & You are a macro-level GIS expert. Your task is to identify, from the POI hidden representation, only the highest-level administrative division (province, autonomous region, or municipality). Ignore finer-grained city or town information and focus solely on determining the provincial-level jurisdiction. \textit{Output format: province name (output ``Unknown'' if not determinable).} Hidden representation: \\[4pt]
T2 & You are a provincial-level GIS expert. Your task is to precisely identify the prefecture-level city (or autonomous prefecture) to which the POI belongs, based on the hidden representation together with the province information already recovered. Ensure that the identified city genuinely belongs to the given province. \textit{Output format: city name (output ``Unknown'' if not determinable).} Hidden representation: \\[4pt]
T3 & You are a local GIS expert. Your task is to identify the district, county, or county-level city to which the POI belongs, based on the hidden representation together with the province and city information already recovered. Ensure the identified district lies within the given city. \textit{Output format: district name (output ``Unknown'' if not determinable).} Hidden representation: \\[4pt]
T4 & You are a micro-level navigation assistant. Your task is to precisely locate the specific township, street, or road based on the POI hidden representation together with the province, city, and district information already recovered. Use the known administrative context as background knowledge to extract fine-grained location descriptions from the hidden representation. \textit{Output format: township or street name (output ``Unknown'' if not determinable).} Hidden representation: \\
\bottomrule
\end{tabular}
\vspace{-0.3cm}
\end{table*}

\begin{table*}[t]
\centering
\small
\caption{Semantic stream PGD instruction prompts (T5--T7).}
\vspace{-0.3cm}
\label{tab:pgd_prompts_entity}
\setlength{\tabcolsep}{4pt}
\begin{tabular}{p{0.06\textwidth} p{0.88\textwidth}}
\toprule
\textbf{Task} & \textbf{Instruction prompt} \\
\midrule
T5 & You are a professional POI entity recognition expert. Your task is to decode the POI hidden representation into standardized store core information to obtain the accurate brand name of the store. \textit{Output format: brand name (brand name must be specific, e.g., KFC, MIXUE; output ``Unknown'' if not determinable).} Hidden representation: \\[4pt]
T6 & You are a professional POI entity recognition expert. Your task is to analyze the primary business category of the store from the POI hidden representation; accuracy is required. \textit{Output format: primary category (must be specific, e.g., ``local specialty snacks'' is preferred over ``dining'').} Hidden representation: \\[4pt]
T7 & You are a seasoned local-life store exploration expert. Your task is to identify detailed store information from the POI hidden representation. \textit{Requirement: naturally describe best-selling products, featured combos, and suitable customers; keep the length moderate without piling up adjectives.} Hidden representation: \\
\bottomrule
\end{tabular}
\vspace{-0.3cm}
\end{table*}

\section{Additional Experimental Results}
\label{app:additional_experiments}

\subsection{Joint Encoding versus Field-wise Structured Encoding}
\label{app:structured_field_baseline}

A natural alternative to LGRID is to encode different item-side
descriptions independently. We refer to this baseline as field-wise
structured encoding. It separately encodes geographic descriptions,
merchant information, and category information, thereby preserving field
boundaries but preventing cross-field interaction during encoding. LGRID
instead first encodes the item as a unified textual representation and
then applies structural decoupling to organize the mixed hidden states
into attribute-aligned slots.

\textbf{Attribute-consistent retrieval protocol.}
For each query POI, we retrieve top-ranked POIs from the full corpus using
the learned representation and evaluate whether the retrieved POIs share
the target attribute with the query. We report MRR and Recall@$K$ at
$K=\{1,5,10,50,100\}$. Higher values indicate stronger attribute-consistent retrieval.
Table~\ref{tab:structured_field_baseline} shows that LGRID consistently
outperforms field-wise encoding on all six attributes. The gains are
larger on fine-grained locations: District and Town MRR improve by
$45.7\%$ and $43.3\%$, and their Recall@100 gains reach $99.5\%$ and
$95.2\%$. Semantic attributes also improve, with Brand and Category MRR
increasing by $60.9\%$ and $88.5\%$. These results show that independent
field encoding preserves field identities but misses cross-field geo--semantic interactions, which are important for fine-grained
disambiguation.
These results support LGRID’s design: joint encoding captures geo--content interactions, while structural decoupling organizes shared representations into interpretable slots. Field-wise encoding misses cross-field interactions, whereas joint encoding alone leaves SIDs entangled.

\begin{table}[t]
\centering
\small
\caption{Joint encoding versus field-wise structured encoding under
attribute-consistent retrieval.}
\vspace{-0.3cm}
\label{tab:structured_field_baseline}
\setlength{\tabcolsep}{2.6pt}
\renewcommand{\arraystretch}{1.05}
\begin{tabular}{@{}llcccccc@{}}
\toprule
\textbf{Dim.} & \textbf{Method} & \textbf{MRR}
& \textbf{R@1} & \textbf{R@5} & \textbf{R@10}
& \textbf{R@50} & \textbf{R@100} \\
\midrule
Prov.    & Field-wise & 0.751 & 0.717 & 0.698 & 0.683 & 0.636 & 0.611 \\
Prov.    & LGRID      & 0.987 & 0.984 & 0.979 & 0.970 & 0.931 & 0.909 \\
\midrule
City     & Field-wise & 0.719 & 0.667 & 0.634 & 0.609 & 0.526 & 0.482 \\
City     & LGRID      & 0.990 & 0.982 & 0.963 & 0.949 & 0.889 & 0.853 \\
\midrule
District & Field-wise & 0.633 & 0.551 & 0.489 & 0.442 & 0.300 & 0.240 \\
District & LGRID      & 0.922 & 0.864 & 0.809 & 0.765 & 0.579 & 0.479 \\
\midrule
Town     & Field-wise & 0.334 & 0.241 & 0.179 & 0.150 & 0.086 & 0.067 \\
Town     & LGRID      & 0.479 & 0.344 & 0.284 & 0.250 & 0.164 & 0.131 \\
\midrule
Brand    & Field-wise & 0.204 & 0.119 & 0.125 & 0.124 & 0.128 & 0.130 \\
Brand    & LGRID      & 0.328 & 0.196 & 0.205 & 0.212 & 0.228 & 0.241 \\
\midrule
Category & Field-wise & 0.052 & 0.048 & 0.050 & 0.051 & 0.053 & 0.054 \\
Category & LGRID      & 0.098 & 0.083 & 0.086 & 0.088 & 0.092 & 0.095 \\
\bottomrule
\end{tabular}
\vspace{-0.3cm}
\end{table}

\begin{table*}[t]
\centering
\small
\vspace{-0.3cm}
\caption{Extended component-level ablation on Kuaishou. Accuracy measures
LLM-based attribute decoding. R@50 measures attribute-consistent
nearest-neighbor retrieval.}
\label{tab:failure_mode_ablation_full}
\vspace{-0.2cm}
\setlength{\tabcolsep}{3.5pt}
\renewcommand{\arraystretch}{1.05}
\begin{tabular*}{\textwidth}{@{\extracolsep{\fill}}l p{0.26\textwidth} c c c c c c@{}}
\toprule
\textbf{Variant} & \textbf{Removed functionality}
& \textbf{Prov. Acc} & \textbf{Town Acc} & \textbf{Brand Acc}
& \textbf{Dist. R@50} & \textbf{Town R@50} & \textbf{Brand R@50} \\
\midrule
Full LGRID & -- & 99.02 & 92.34 & 86.62 & 0.679 & 0.250 & 0.921 \\
w/o $\mathcal{L}_{\mathrm{PGD}}$ & Generative slot decodability is removed. & 78.40 & 10.70 & 44.80 & 0.452 & 0.121 & 0.841 \\
w/o SCR & Structured routing between slots is removed. & 75.14 & 6.85 & 16.65 & 0.291 & -- & -- \\
w/o $\mathcal{L}_{\mathrm{SACL}}$ & Structure-aware contrastive discrimination is removed. & 93.80 & 34.60 & 48.70 & 0.336 & 0.089 & 0.768 \\
w/o $\mathcal{L}_{\mathrm{LDR}}$ & Slot diversity regularization is removed. & 98.30 & 81.90 & 77.80 & 0.598 & 0.204 & 0.879 \\
SACL-Simple & Only simple negatives are used. & 89.36 & -- & 42.66 & 0.401 & -- & -- \\
SACL-HierHard & Hierarchical hard negatives are used without the full fallback strategy. & 94.52 & -- & 78.95 & 0.561 & -- & -- \\
\bottomrule
\end{tabular*}
\vspace{-0.2cm}
\end{table*}
\vspace{-0.2cm}
\subsection{Codebook Collision and Prefix Utilization}
\label{app:codebook_collision}

To evaluate the distinguishability of generated SIDs, we report the
full-SID collision rate and cumulative prefix utilization, referred to as
$n$-gram utilization in the main text. For LGRID, the full SID of item
$i$ is obtained by serializing the geographic and semantic SIDs:
\begin{equation}
\mathbf{C}_i=[\mathbf{C}^{g}_i;\mathbf{C}^{s}_i].
\end{equation}

Let $\mathbf{C}_i=(c_{i,1},\ldots,c_{i,L})$ denote the serialized full
SID. The full-SID collision rate is:
\vspace{-0.2cm}
\begin{equation}
\mathrm{Collision}
=
1-\frac{|\{\mathbf{C}_i\}_{i=1}^{N}|}{N},
\end{equation}
where $N$ is the number of items. A lower collision rate means that fewer
distinct POIs share the same full SID.
For prefix length $m$, cumulative prefix utilization is:
\begin{equation}
U_{1:m}
=
100\times
\frac{
|\{(c_{i,1},\ldots,c_{i,m})\}_{i=1}^{N}|
}{
\prod_{j=1}^{m}K_j
},
\end{equation}
where $K_j$ is the codebook size at level $j$. This form allows different
levels to have different codebook sizes. Higher utilization indicates
that the model uses a larger portion of the available prefix space.
Because LGRID separates geographic and semantic code streams, the same
collision metric applies to $\mathbf{C}^{g}_i$ and $\mathbf{C}^{s}_i$
separately. This enables stream-wise analysis of whether repeated full
SIDs mainly arise from geographic or semantic codes, which is unavailable
for single-stream SID methods.

\subsection{Extended Component-level Ablation}
\label{app:failure_mode_ablation}

Table~\ref{tab:failure_mode_ablation_full} extends the component-level
ablation in the main text with additional decoding and retrieval metrics.
The goal is to identify which aspect of representation quality is most
affected when each component is removed.
The extended results are consistent with the main-text ablation. Removing
SCR causes the largest degradation across decoding and retrieval metrics,
showing that structured routing is central to slot separation. Removing
$\mathcal{L}_{\mathrm{PGD}}$ mainly hurts slot decodability, especially
for fine-grained Town recovery. Removing $\mathcal{L}_{\mathrm{SACL}}$
substantially weakens discrimination and retrieval, while removing
$\mathcal{L}_{\mathrm{LDR}}$ causes smaller but consistent drops,
supporting its role in reducing redundant slot usage.

\vspace{-0.1cm}
\subsection{Prompt Robustness}
\label{app:prompt_robustness}
Semantic anchors only initialize the slot queries in $\mathbf{Z}^{(0)}$; they are not supervision labels and do not contain
target answers. They place each slot query near the semantic neighborhood
of its intended attribute, such as \textit{Province} or \textit{Brand},
helping cross-attention establish stable slot--attribute alignment.

To verify that this effect does not depend on one specific wording
choice, we keep the SD-Block architecture and training strategy fixed and
only change the anchor phrases used for initialization. We compare the
original anchors with synonym anchors, natural-language anchors, and
random initialization. Table~\ref{tab:prompt_robustness} reports slot-level attribute decoding accuracy under the same interpretability task as
the main-text learning-strategy ablation.

\begin{table}[t]
\centering
\small
\caption{Robustness of semantic anchor choices on attribute decoding accuracy (\%).}
\vspace{-0.3cm}
\label{tab:prompt_robustness}
\setlength{\tabcolsep}{3.5pt}
\renewcommand{\arraystretch}{1.05}
\begin{tabular}{lcccccc}
\toprule
\textbf{Method} & \textbf{Prov.} & \textbf{City} & \textbf{Dist.} &
\textbf{Town} & \textbf{Brand} & \textbf{Cat.} \\
\midrule
Random Init & 46.4 & 34.6 & 22.2 & 4.51 & 31.9 & 37.4 \\
Original Anchor & 99.0 & 99.7 & 99.1 & 92.3 & 86.6 & 97.0 \\
Synonym Anchor & 98.2 & 98.8 & 98.2 & 91.4 & 84.8 & 96.2 \\
Natural-language Anchor & 97.6 & 98.3 & 98.0 & 90.7 & 85.1 & 95.8 \\
\bottomrule
\end{tabular}
\vspace{-0.3cm}
\end{table}
The three semantically meaningful anchor variants achieve similar
accuracy, whereas random initialization sharply degrades decoding,
especially on the fine-grained \textit{Town} attribute. This shows that
semantic anchors stabilize slot--attribute alignment rather than acting as
a hidden prompt trick or answer injection.

\begin{table*}[t]
\centering
\vspace{-0.3cm}
\caption{Full attribute-consistent retrieval results on Kuaishou. We report
Recall (Rec), Coverage (Cov), and Hit Rate (Hit) at multiple cutoffs. The baseline is the entangled LLM representation without LGRID's
structured slot decoupling.}
\label{tab:full_recall}
\vspace{-0.3cm}
\small
\setlength{\tabcolsep}{1.1pt} 
\renewcommand{\arraystretch}{1.1} 
\resizebox{\textwidth}{!}{%
\begin{tabular}{llc ccc ccc ccc ccc ccc}
\toprule
\multirow{2}{*}{Dim.} & \multirow{2}{*}{Model} & \multirow{2}{*}{MRR} &
\multicolumn{3}{c}{@1} & \multicolumn{3}{c}{@5} & \multicolumn{3}{c}{@10} &
\multicolumn{3}{c}{@50} & \multicolumn{3}{c}{@100} \\
\cmidrule(lr){4-6}\cmidrule(lr){7-9}\cmidrule(lr){10-12}\cmidrule(lr){13-15}\cmidrule(lr){16-18}
& & & Rec & Cov & Hit & Rec & Cov & Hit & Rec & Cov & Hit & Rec & Cov & Hit & Rec & Cov & Hit \\
\midrule

\multirow{3}{*}{Prov.}
& Base & 0.9373 & 0.924 & 0.924 & 0.924 & 0.902 & 0.902 & 0.963 & 0.881 & 0.881 & 0.988 & 0.750 & 0.750 & 0.998 & 0.668 & 0.668 & 0.999 \\
& \textbf{Ours} & \textbf{0.9592} & \textbf{0.943} & \textbf{0.943} & \textbf{0.943} & \textbf{0.918} & \textbf{0.918} & \textbf{0.980} & \textbf{0.903} & \textbf{0.903} & \textbf{0.990} & \textbf{0.854} & \textbf{0.854} & \textbf{0.998} & \textbf{0.825} & \textbf{0.825} & \textbf{0.999} \\
&  \cellcolor{lightgray}\textbf{Gain} & \imp{+2.3\%} & \imp{+2.1\%} & \imp{+2.1\%} & \imp{+2.1\%} & \imp{+1.8\%} & \imp{+1.8\%} & \imp{+1.8\%} & \imp{+2.5\%} & \imp{+2.5\%} & \imp{+0.2\%} & \imp{+13.9\%} & \imp{+13.9\%} & \imp{0.0\%} & \imp{+23.5\%} & \imp{+23.5\%} & \imp{0.0\%} \\
\midrule

\multirow{3}{*}{City}
& Base & 0.9150 & 0.892 & 0.892 & 0.892 & 0.807 & 0.807 & 0.942 & 0.745 & 0.745 & 0.956 & 0.543 & 0.543 & 0.979 & 0.443 & 0.443 & 0.987 \\
& \textbf{Ours} & \textbf{0.9443} & \textbf{0.925} & \textbf{0.925} & \textbf{0.925} & \textbf{0.896} & \textbf{0.896} & \textbf{0.967} & \textbf{0.877} & \textbf{0.877} & \textbf{0.977} & \textbf{0.819} & \textbf{0.819} & \textbf{0.991} & \textbf{0.786} & \textbf{0.786} & \textbf{0.994} \\
& Gain & \imp{+3.2\%} & \imp{+3.7\%} & \imp{+3.7\%} & \imp{+3.7\%} & \imp{+11.0\%} & \imp{+11.0\%} & \imp{+2.7\%} & \imp{+17.7\%} & \imp{+17.7\%} & \imp{+2.2\%} & \imp{+50.8\%} & \imp{+50.8\%} & \imp{+1.2\%} & \imp{+77.4\%} & \imp{+77.4\%} & \imp{+0.7\%} \\
\midrule

\multirow{3}{*}{Dist.}
& Base & 0.7402 & 0.689 & 0.689 & 0.689 & 0.537 & 0.539 & 0.801 & 0.450 & 0.452 & 0.834 & 0.249 & 0.252 & 0.891 & 0.181 & 0.184 & 0.911 \\
& \textbf{Ours} & \textbf{0.8832} & \textbf{0.842} & \textbf{0.842} & \textbf{0.842} & \textbf{0.794} & \textbf{0.795} & \textbf{0.934} & \textbf{0.765} & \textbf{0.766} & \textbf{0.952} & \textbf{0.679} & \textbf{0.680} & \textbf{0.976} & \textbf{0.631} & \textbf{0.632} & \textbf{0.984} \\

& \cellcolor{lightgray}\textbf{Gain} & \imp{+19.3\%} & \imp{+22.2\%} & \imp{+22.2\%} & \imp{+22.2\%} & \imp{+47.9\%} & \imp{+47.5\%} & \imp{+16.6\%} & \imp{+70.0\%} & \imp{+69.5\%} & \imp{+14.1\%} & \imp{+172.7\%} & \imp{+169.8\%} & \imp{+9.5\%} & \imp{+248.6\%} & \imp{+243.5\%} & \imp{+8.0\%} \\
\midrule

\multirow{3}{*}{Town}
& Base & 0.4284 & 0.356 & 0.356 & 0.356 & 0.229 & 0.240 & 0.511 & 0.176 & 0.186 & 0.569 & 0.087 & 0.090 & 0.679 & 0.064 & 0.062 & 0.720 \\
& \textbf{Ours} & \textbf{0.5137} & \textbf{0.391} & \textbf{0.391} & \textbf{0.391} & \textbf{0.342} & \textbf{0.402} & \textbf{0.661} & \textbf{0.315} & \textbf{0.373} & \textbf{0.749} & \textbf{0.250} & \textbf{0.290} & \textbf{0.884} & \textbf{0.228} & \textbf{0.250} & \textbf{0.917} \\
& \cellcolor{lightgray}\textbf{Gain} & \imp{+19.9\%} & \imp{+9.8\%} & \imp{+9.8\%} & \imp{+9.8\%} & \imp{+49.3\%} & \imp{+67.5\%} & \imp{+29.4\%} & \imp{+79.0\%} & \imp{+100.5\%} & \imp{+31.6\%} & \imp{+187.4\%} & \imp{+222.2\%} & \imp{+30.2\%} & \imp{+256.3\%} & \imp{+303.2\%} & \imp{+27.4\%} \\
\midrule

\multirow{3}{*}{Brand}
& Base & 0.9532 & 0.936 & 0.936 & 0.936 & 0.929 & 0.925 & 0.972 & 0.923 & 0.913 & 0.978 & 0.902 & 0.870 & 0.994 & 0.894 & 0.842 & 0.996 \\
& \textbf{Ours} & \textbf{0.9713} & \textbf{0.958} & \textbf{0.958} & \textbf{0.958} & \textbf{0.946} & \textbf{0.941} & \textbf{0.988} & \textbf{0.939} & \textbf{0.930} & \textbf{0.993} & \textbf{0.921} & \textbf{0.890} & \textbf{0.997} & \textbf{0.914} & \textbf{0.863} & \textbf{0.998} \\
& \cellcolor{lightgray}\textbf{Gain} & \imp{+1.9\%} & \imp{+2.4\%} & \imp{+2.4\%} & \imp{+2.4\%} & \imp{+1.8\%} & \imp{+1.7\%} & \imp{+1.6\%} & \imp{+1.7\%} & \imp{+1.9\%} & \imp{+1.5\%} & \imp{+2.1\%} & \imp{+2.3\%} & \imp{+0.3\%} & \imp{+2.2\%} & \imp{+2.5\%} & \imp{+0.2\%} \\

\midrule

\multirow{3}{*}{Cat.}
& Base & 0.9179 & 0.893 & 0.893 & 0.893 & 0.879 & 0.879 & 0.968 & 0.854 & 0.854 & 0.976 & 0.838 & 0.838 & 0.994 & 0.817 & 0.817 & 0.997 \\
& \textbf{Ours} & \textbf{0.9363} & \textbf{0.906} & \textbf{0.906} & \textbf{0.906} & \textbf{0.891} & \textbf{0.891} & \textbf{0.974} & \textbf{0.882} & \textbf{0.882} & \textbf{0.985} & \textbf{0.855} & \textbf{0.855} & \textbf{0.996} & \textbf{0.839} & \textbf{0.839} & \textbf{0.998} \\

& \cellcolor{lightgray}\textbf{Gain} & \imp{+2.0\%} & \imp{+1.5\%} & \imp{+1.5\%} & \imp{+1.5\%} & \imp{+1.4\%} & \imp{+1.4\%} & \imp{+0.6\%} & \imp{+3.3\%} & \imp{+3.3\%} & \imp{+0.9\%} & \imp{+2.0\%} & \imp{+2.0\%} & \imp{+0.2\%} & \imp{+2.7\%} & \imp{+2.7\%} & \imp{+0.1\%} \\

\bottomrule
\end{tabular}
}
\end{table*}

\begin{figure*}[t]
  \centering
  \includegraphics[width=0.9\textwidth]{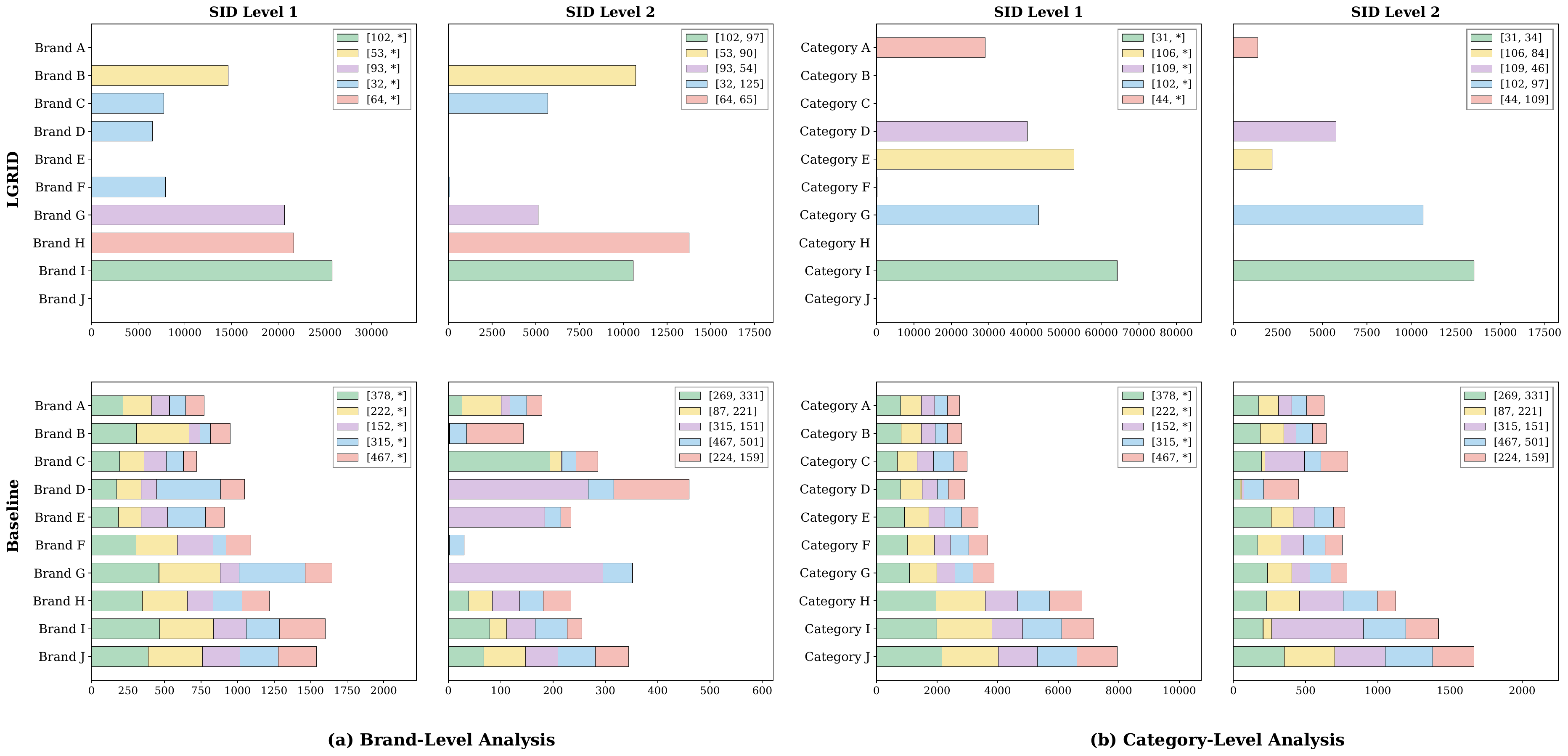}
  \vspace{-0.4cm}
  \caption{Brand- and category-level distributions of top-frequency SID prefixes.}
  \label{fig:ap-bar}
  \vspace{-0.3cm}
\end{figure*}

\subsection{Full Attribute-consistent Retrieval Results}
\label{app:full_retrieval}

This section reports the full attribute-consistent retrieval results that
complement Table~\ref{tab:recall_res} in the main text. We evaluate
retrieval over the full POI corpus, covering both geographic attributes
(\textit{Province}, \textit{City}, \textit{District}, and \textit{Town})
and semantic attributes (\textit{Brand} and \textit{Category}). The
baseline denotes the corresponding entangled LLM representation without
LGRID's structured slot decoupling.

For a query POI, let $\mathcal{R}_K(q)$ be the top-$K$ retrieved POIs and
let $\mathcal{G}(q)$ be the set of POIs sharing the target attribute with
$q$. We report four metrics:
\begin{itemize}
    \item \textbf{Recall@$K$}: the number of matched retrieved POIs
    normalized by $\min(K,|\mathcal{G}(q)|)$.
    \item \textbf{Coverage@$K$}: the fraction of top-$K$ retrieved POIs
    that match the target attribute.
    \item \textbf{Hit@$K$}: whether at least one retrieved POI matches the
    target attribute.
    \item \textbf{MRR}: the reciprocal rank of the first matched POI.
\end{itemize}
Queries with missing target attributes are excluded from the
corresponding attribute dimension.

Table~\ref{tab:full_recall} shows that LGRID improves retrieval
consistency across geographic and semantic attributes. Gains are modest
for coarse attributes such as Province, where the baseline is strong, but
grow at finer geographic levels. District MRR rises from $0.7402$ to
$0.8832$ ($+19.3\%$), and Town MRR from $0.4284$ to $0.5137$
($+19.9\%$). Larger cutoffs show stronger gains: District Recall@50
improves by $172.7\%$, and Town Recall@100 by $256.3\%$. These results
suggest that LGRID preserves fine-grained geographic consistency while
maintaining stable Brand and Category retrieval.

\vspace{-0.1cm}
\subsection{Additional SID-prefix Visualizations}
\label{app:additional_sid_visualization}

Fig.~\ref{fig:ap-bar} extends the SID-prefix analysis to semantic
attributes. It visualizes the distribution of top-frequency SID prefixes
for Brand and Category. LGRID produces more concentrated prefix
distributions, indicating that semantically related POIs are more
consistently assigned to shared or nearby prefix patterns. In contrast,
the baseline spreads the same semantic attributes across many prefixes,
suggesting weaker semantic discrimination and a higher risk of prefix
collapse. This visualization complements the city/district analysis in
the main text and shows that LGRID's structured SID usage is not limited
to geographic attributes.

\end{appendices}
\end{document}